\documentclass{sig-alternate-05-2015}
\usepackage{graphicx}
\usepackage{balance}  
\usepackage{listings}
\usepackage{pslatex}
\usepackage{amsmath}
\usepackage{amssymb}
\usepackage{graphicx}
\usepackage{flushend}
\usepackage{xcolor}
\usepackage{url}
\usepackage{listings}
\usepackage{paralist}
\usepackage{xspace}
\usepackage{epstopdf}
\usepackage{tablefootnote}
\usepackage[skip=6pt]{caption, subcaption}
\usepackage{xifthen}
\usepackage{balance}
\usepackage{tikz}
\usetikzlibrary{shapes,arrows}
\usepackage{pgfplots}
\usepackage{capt-of}
\usepackage[T1]{fontenc}
\usepackage[scaled]{beramono}
\usepackage{booktabs}
\usepackage{array}
\usepackage{ragged2e}
\usepackage{csvsimple}
\usepackage{calc}
\usepackage[english]{babel}
\usepackage{blindtext}

\usepackage{etoolbox}
\makeatletter
\patchcmd{\maketitle}{\@copyrightspace}{}{}{}
\makeatother

\begin{document}

\title{Push vs. Pull-Based Loop Fusion in Query Engines}



\author{
\alignauthor 
Amir Shaikhha, Mohammad Dashti, and Christoph Koch
\hspace{1cm}\\
\affaddr{
\{firstname\}.\{lastname\}@epfl.ch}\\ 
\affaddr{\'{E}cole Polytechnique F\'{e}d\'{e}rale de Lausanne}
}

\maketitle

\begin{abstract}
Database query engines use pull-based or push-based approaches to avoid the materialization of data across query operators. In this paper, we study these two types of  query engines in depth and present the limitations and advantages of each engine. Similarly, the programming languages community has developed loop fusion techniques to remove intermediate collections in the context of collection programming.  We draw parallels between the DB and PL communities by demonstrating the connection between pipelined query engines and loop fusion techniques. Based on this connection, we propose a new type of pull-based engine, inspired by a loop fusion technique, which combines the benefits of both approaches. Then we experimentally evaluate the various engines, in the context of query compilation, for the first time in a fair environment, eliminating the biasing impact of ancillary optimizations that have traditionally only been used with one of the approaches. 
We show that for realistic analytical workloads, there is no considerable advantage for either form of pipelined query engine, as opposed to what recent research suggests.
Also, by using microbenchmarks we show that our proposed engine dominates the existing engines by combining the benefits of both. 
\end{abstract}


\lstdefinelanguage{Scala}%
{morekeywords={abstract,%
  case,catch,char,class,%
  def,else,extends,final,finally,for,%
  if,import,implicit,%
  match,module,%
  new,null,%
  object,override,%
  package,private,protected,public,%
  for,public,return,super,%
  this,throw,trait,try,type,%
  val,var,%
  with,while,do,%
  yield, sealed
  },%
  sensitive,%
  morecomment=[l]//,%
  morecomment=[s]{/*}{*/},%
  morestring=[b]",%
  morestring=[b]',%
  showstringspaces=false%
}[keywords,comments,strings]%

\lstset{language=Scala,%
  escapeinside={(*@}{@*)},%
  breaklines=true,%
  mathescape=true,%
showspaces=false,
showtabs=false,
showstringspaces=false,
breakatwhitespace=true,
  aboveskip=1pt,
  belowskip=1pt,
  lineskip=-0.2pt,
   numbersep=5pt,
   numberstyle=\tiny\ttfamily,
   basicstyle=\scriptsize\ttfamily,
  keywordstyle=\scriptsize\ttfamily\bfseries,%
columns=fullflexible,
  emph={%
    build, destroy,%
  next, consume, generate,
  stream, unstream, foreach
    },emphstyle={\color{blue}\bfseries},
  escapeinside={(*@}{@*)}
}

\definecolor{listingbg}{RGB}{240, 240, 240}
\newcommand{\code}[1]{\lstinline[language=Scala,columns=fixed,basicstyle=\ttfamily,keywordstyle=\ttfamily,emph={}]|#1|}
\newcommand{\codenorm}[1]{\lstinline[language=Scala,columns=fixed]|#1|}
\newcommand{\systemname}{DBLAB/LB\xspace}
\newcommand{\dblab}{DBLAB\xspace}


\newcommand{\note}[1]{{\color{red}[#1]}}
\newcommand{\todo}[1]{\note{TODO: #1}}
\newcommand{\cktodo}[1]
{\note{CKTODO: #1}}

\def\tpch{\mbox{TPC-H}\xspace}
\def\timesten{\mbox{DBX}\xspace}
\def\clang{\mbox{CLang}\xspace}
\def\glib{\mbox{GLib}\xspace}
\def\legobase{\mbox{LegoBase}\xspace}
\def\legobaseorig{\mbox{LegoBase}\xspace}
\def\pardis{\mbox{SC}\xspace}
\def\groupby{\mbox{group by}\xspace}
\def\datastructure{data-stru\-cture\xspace}
\def\datastructures{data-stru\-ctures\xspace}
\def\hyper{HyPer\xspace}
\def\gcc{GCC\xspace}
\def\dblab{DBLAB\xspace}

\newcommand\christoph[1]{{\textcolor{red}{[CK: #1]}}}
\newcommand\lionel[1]{{\textcolor{blue}{[LP: #1]}}}

\def\sigmaspl{$\Sigma$-SPL}
\def\sigmall{$\Sigma$-LL}

\def\multimap{\texttt{MultiMap}}
\def\hashmap{\texttt{HashMap}}
\def\array{\texttt{Array}}
\def\arraybuffer{\texttt{ArrayBuffer}}
\def\treeset{\texttt{TreeSet}}

\def\Collection{List}
\def\collection{list}
\def\collectionType{List}
\def\collectionCode{\code{List}}

\def\prossign{{\color[rgb]{0,0.5,0}\checkmark}}
\def\conssign{{\color{red}\ding{55}}}

\definecolor{nice_blue}{rgb}{0.45,0.82,0.92}
\definecolor{nice_red}{rgb}{0.86,0.30, 0.28}

\def\hashtable{hash table}
\def\Hashtable{Hash table}
\def\smartpull{inline-aware}
\def\Smartpull{Inline-aware}
\def\SmartPull{Inline-Aware}


\newcommand{\dsl}[1]{#1}

\def\qopl{\dsl{QPlan}}
\def\qml{\dsl{QMonad}}
\def\mchl{\dsl{ScaLite[Map, \collectionType{}]}}
\def\mcl{\dsl{ScaLite[\collectionType{}]}}
\def\scalaql{\dsl{ScaLite}}
\def\cscalaql{\dsl{C.Scala}}

\def\cql{\dsl{C}}

\newcommand{\para}[1]{\vspace{0.1cm}\noindent\textbf{#1}.}
\newcommand{\fusionrule}[3]{\vspace{0.1cm} \noindent \textit{\textbf{#1}}: \\ \code{#2} \hspace{0.1cm} $ \leadsto$ \hspace{0.1cm} \code{#3} \vspace{0.1cm}}

\def\figurebottom{\vspace{-0.2cm}}



\definecolor{col1}{rgb}{0.8,0.9,1}
\definecolor{col2}{rgb}{0.8,0.9,1}

\newlength{\DepthReference}
\setlength{\DepthReference}{0.5pt}

\newlength{\HeightReference}
\setlength{\HeightReference}{2.5pt}

\newlength{\Width}%
\newcommand{\MyColorBox}[2][col1]%
{%
    \settowidth{\Width}{#2}%
    \colorbox{#1}%
    {%
        \raisebox{-\DepthReference}%
        {%
                \parbox[b][\HeightReference+\DepthReference][c]{\Width}{\centering#2}%
        }%
    }%
}

\section{Introduction}
\label{sec:intro}
Database query engines successfully leverage the compositionality of relational algebra-style query plan languages. Query plans are compositions of operators that, at least conceptually, can be executed in sequence, one after the other. However, actually evaluating queries in this way leads to grossly suboptimal performance. Computing (``materialising'') the result of a first operator before passing it to a second operator can be very expensive, particularly if the intermediate result is large and needs to be pushed down the memory hierarchy. The same observation has been made by the programming languages and compilers community and has led to work on loop fusion and deforestation (the elimination of data structure construction and destruction for intermediate results).

Already relatively early on in the history of relational database systems, a solution to this problem has been proposed in the form of the Volcano Iterator model~\cite{Volcano}. In this model, tuples are {\em pulled}\/ up through a chain of operators that are linked by iterators that advance in lock-step. Intermediate results between operators are not accumulated, but tuples are produced on demand, by request by conceptually ``later'' operators.

More recently, an operator chaining model has been proposed that shares the advantage of avoiding materialisation of intermediate results but which reverses the control flow; tuples are {\em pushed}\/ forward from the source relations to the operator producing the final result. Recent papers~\cite{Neumann11, legobase} seem to suggest that this push-model consistently leads to better query processing performance than the pull model, even though no direct, fair comparisons are provided.

One of the main contributions of this paper is to debunk this myth. As we show, if compared fairly, push and pull based engines have very similar performance, with individual strengths and weaknesses, and neither is a clear winner. Push engines have in essence only been considered in the context of query compilation, conflating the potential advantages of the push paradigm with those of code inlining. To compare them fairly, one has to decouple these aspects.

\begin{figure}[t]
\includegraphics[width=\columnwidth, height=2.7cm]{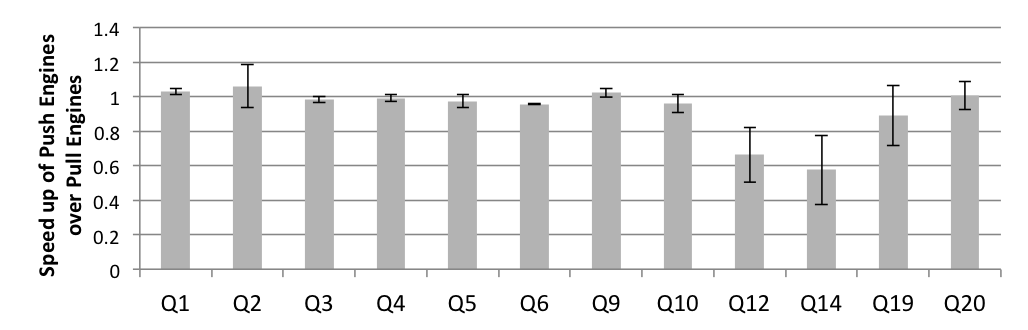}
\caption{Performance comparison of push-based and pull-based query engines using TPCH queries running 8 GBs of data.}
\figurebottom
\vspace{-0.3cm}
\label{fig:tizer}
\end{figure}

Figure~\ref{fig:tizer} shows a performance comparison of these two engines for several \tpch queries using 8 GBs of data in a fair scenario. There is no clear winner among these two engines. In the case of two queries (\tpch queries 12 and 14), the pull engine is performing better than the push engine. However, in some cases, the push-based query engine is performing marginally better. The advantages and limitations of these engines are explained in more detail in Section~\ref{sec:engine}.

In this paper, we present an in-depth study of the tradeoffs of the push versus the pull paradigm.  Choosing among push and pull -- or any resonable alternative -- is a fundamental decision which drives many decisions throughout the architecture of a query engine. Thus, one must understand the relevant properties and tradeoffs deeply, and should not bet on one's ability to overcome the disadvantages of a choice by a hack later.

Furthermore, we illustrate how the same challenge and tradeoff has been met and addressed by the PL community, and show a number of results that can be carried over from the lessons learned there. Specifically, we study how the PL community's answer to the problem, {\em stream fusion}\/~\cite{Coutts07streamfusion}, can be adapted to the query processing scenario, and show how it combines the advantages of the pull and push approaches. Furthermore, we demonstrate how we can use ideas from the push approach to solve well-known limitations of stream fusion. As a result, we construct a query engine which combines the benefits of both push and pull approaches. In essence, this engine is a pull-based engine on a coarse level of granularity, however, on a finer level of granularity, it pushes the individual data tuples.

In summary, this paper makes the following contributions:
\begin{itemize}
\item We discuss pipelined query engines in Section~\ref{sec:engine}. After presenting loop fusion for collection programming in Section~\ref{sec:fusion}, we show the connection between these two concepts in Section~\ref{sec:cor}. Furthermore, we demonstrate the limitations associated to each approach.
\item Based on this connection with loop fusion, we propose a new pipelined query engine in Section~\ref{sec:stream} inspired by the stream fusion~\cite{Coutts07streamfusion} technique developed for collection programming in the PL community.
Also, we discuss implementation concerns and compiler optimizations required for the proposed pipelined query engine in Section~\ref{sec:impl}.
\item We experimentally evaluate the various query engine architectures in Section~\ref{sec:exp}. Using microbenchmarks, we discuss the weaknesses of the existing engines and how the proposed engine circumvents these weaknesses by combining the benefits of both worlds. Then we demonstrate using \tpch queries that good implementations of these engines do not show a considerable advantage for either form of pipelined query engine.
\end{itemize}

Throughout this paper, we are using the Scala programming language for all code snippets, interfaces and examples. None of the concepts and ideas require specifically this language -- other impure functional object-oriented programming languages such as OCaml, F\#, C++11, C\#, or Java 8 could be used instead.

\section{Pipelined Query Engines}
\label{sec:engine}
Database management systems accept a declarative query (e.g., written in SQL). Such a query is passed to a query optimizer to find a fast physical query plan, which then is either interpreted by the query engine or compiled to low-level code (e.g. C code).

Physical query plans perform calculations and data transformations. A sequence of query operators can be \textit{pipelined}, which means that the output of one operator is {\em streamed} into the next operator without materializing the intermediate data.

There are two approaches for pipelining. The first approach is demand-driven pipelining in which an operator repeatedly \textit{pulls} the next data tuple from its \textit{source} operator. The second approach is data-driven pipelining in which an operator \textit{pushes} each data tuple to its \textit{destination} operator.
Next, we give more details on the pull-based and push-based query engines.

\subsection{Pull Engine -- a.k.a.\ the Iterator Pattern}
\label{sec:engine-pull}
The iterator model is the most widely used pipelining technique in query engines. This model was initially proposed in XRM~\cite{lorie1974xrm}. However, the popularity of this model is due to its adoption in the Volcano system~\cite{Volcano}, in which this model was enriched with facilities for parallelization.

In a nutshell, in the iterator model, each operator pipelines the data by requesting the next element from its source operator. This way, instead of waiting until the whole intermediate relation is produced, the data is \textit{lazily} generated in each operator. This is achieved by invoking the \code{next} method of the source operator by the destination operator. The design of pull-based engines directly corresponds to the iterator design pattern in object-oriented programming~\cite{vlissides1995design}.

\begin{figure}[t]
\def\figcomplheightup{2.7cm}
\def\figcomplheightin{.8cm}
\begin{minipage}[c][\figcomplheightup][t]{.73\columnwidth}
\includegraphics[width=\columnwidth]{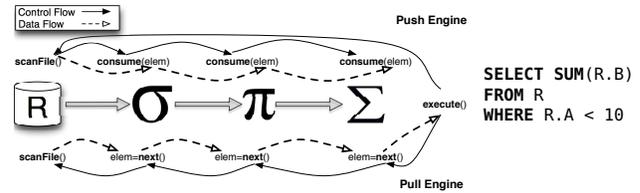}
\end{minipage}
\begin{minipage}[c][\figcomplheightup][t]{.25\columnwidth}
\vspace{.8cm}
\begin{lstlisting}[language=SQL]
SELECT SUM(R.B)
FROM R 
WHERE R.A < 10
\end{lstlisting}
\end{minipage}
\caption{Data flow and control flow for push and pull-based query engine for the provided SQL query.}
\figurebottom
\label{fig:pushpull-example}
\end{figure}

Figure~\ref{fig:pushpull-example} shows an example query and the control flow of query processing for this query. Each query operator performs the role of a destination operator and \textit{requests} data from its source operator (the predecessor operator along the flow direction of data). In a pull engine, this is achieved by invoking the \code{next} function of the source operator, and is shown as control flow edges. In addition, each operator serves as source operator and \textit{generates} result data for its destination operator (the successor operator along the flow direction of data). The generated data is the return value of the \code{next} function, and is represented by the data flow edges in Figure~\ref{fig:pushpull-example}. Note the opposing directions of control-flow and data-flow edges for the pull engine in Figure~\ref{fig:pushpull-example}.

From a different point of view, each operator can be considered as a \code{while} loop in which the \code{next} function of the source operator is invoked per iteration. The loop is terminated when the \code{next} function returns a special value (e.g., a \code{null} value). In other words, whenever this special value is observed, a \code{break} statement is executed to terminate the loop execution.

There are two main issues with a pull-based query engine. First, the \code{next} function invocations are implemented as virtual functions -- operators with different implementations of \code{next} have to be chained together. There are many invocations of these functions, and each of invocation requires looking up a virtual table, which leads to bad instruction locality. Query compilation solves this issue by inlining these virtual function calls, which is explained in Section~\ref{sec:compiled-engine}.

Second, although a pull engine pipelines the data through pipelining operators, in practice, selection operators are problematic. When the \code{next} method of a selection operator is invoked, the destination operator should wait until the selection operator returns the next data tuple satisfying its predicate. This makes the control flow of the query engine more complicated by introducing more loops and branches, which is demonstrated in Figure~\ref{fig:pull-inlined-cfg}. This complicated control flow graph degrades branch prediction. Intuitively, this is because there is no construct for skipping the irrelevant results (such as the \code{continue} construct). This problem is solved in push-based query engines.

\subsection{Push Engine -- a.k.a.\ the Visitor Pattern}
\label{sec:engine_push}

Push-based engines are widely used in streaming systems~\cite{Hirzel:2014:CSP:2597757.2528412}.  The Volcano system uses data-driven pipelining (which is a push-based approach) for implementing inter-operator parallelism in query engines. In the context of query compilation, stream processing engines such as StreamBase \cite{streambase} and Spade \cite{gedik-sigmod:08}, as well as HyPer~\cite{Neumann11} and LegoBase \cite{legobase} use a push-based query engine approach.

In push-based query engines, the control flow is reversed compared to that of pull-based engines. More concretely, instead of destination operators requesting data from their source operators, data is pushed from the source operators towards the destination operators. This is achieved by the source operator passing the data as an argument to the \code{consume} method of the destination operator. This results in \textit{eagerly} transferring the data tuple-by-tuple instead of requesting it \textit{lazily} in pull-engines.  

A push engine can be implemented using the \textit{Visitor} design pattern~\cite{vlissides1995design} from object-oriented programming. This design pattern allows separating an algorithm from a particular type of data. In the case of query engines, the visitor pattern allows us to separate the query operators (data processing algorithms) from a relation of elements. To do so, each operator should be defined as a visitor class, in which the \code{consume} method has the functionality of the \code{visit} method. The process of the initialization of the chain of operators is performed by using the \code{accept} method of the Visitor pattern, which corresponds to the \code{produce} method in push engines.

Figure~\ref{fig:pushpull-example} shows the query processing workflow for the given example query. Query processing in each operator consists of two main phases. In the first phase, operators prepare themselves for producing their data. This is performed only once in the initialization. In the second phase, they consume the data provided by the source operator and produce data for the destination operator. This is the main processing phase, which consists of invoking the \code{consume} method of the destination operator and passing the produced data through it. This results in the same direction for both control-flow and data-flow edges, as shown in Figure~\ref{fig:pushpull-example}.

Push engines solve the problem  pull engines have with selection operators. This is achieved by ignoring the produced data if it does not satisfy the given predicate by using a construct which allows to skip the current iteration of the loop (e.g., using \code{continue}). This simplifies the control flow and improves branch prediction in the case of selection operators. This is in contrast with pull-engines in which the destination operator should have waited for the source operator to serve the request. 

However, push engines experience difficulties with {\em limit} and {\em merge join} operators. For limit operators, push engines do not allow terminating the iteration by nature. This is because, in push engines, the operators cannot control when the data should no longer be produced by their source operator. This causes the production of elements which will never be used.

The merge join operator suffers from a similar problem. There is no way for the merge join operator to guide which one of its two source operators (which are both sorted and there is a 1-to-n relationship between them) should produce the next data item. Hence, it is not possible to pipeline the data coming from both source operators in merge join. As a result, at least for one of the source operators, the pipeline needs to be broken. Hence, the incoming data coming from one of the source operators can be pipelined (assuming it is correctly sorted, of course), but the input data coming from the other source operator must be materialized.

The mentioned limitation is not specific to operators such as merge joins. A similar situation can happen in the case of more sophisticated analytical tasks where one has to use collection programming APIs (such as Spark RDDs~\cite{rdd}). The \code{zip} method in collection programming has a similar behavior to the merge join operator and cannot be easily pipelined in push-based engines.

Note that these limitations can be resolved by providing special cases for these two operators in the push engine. In the case of limit, one can avoid producing unnecessary elements by manually fusing this operator with its source operator (which in most cases is an ordering operator). Also, one can implement a variant of merge join which uses different threads for its source operators and uses synchronization constructs in order to control the production of data by its two inputs, which can be costly. However, in this paper, by push engine, we mean a \textit{pure} push engine without such augmentations.

\subsection{Compiled Engines}
\label{sec:compiled-engine}
In general, the runtime cost of a given query is dependent on two factors. The first factor is the time it takes to transfer the data across storage and computing components. The second factor is the time taken for performing the actual computation (i.e., running the instructions of the query). In disk-based DBMSes, the dominating cost is usually the data transfer from/to the secondary storage. Hence, as long as the pipelining algorithm does not break the pipeline, there is no difference between pull engines and push engines. As a result, the practical problem with selections in pull engines (c.f. Section~\ref{sec:engine-pull}) is obscured by data transfer costs. 


\begin{figure}[t]
\def\figcomplheightup{5.4cm}
\begin{minipage}[c][\figcomplheightup][t]{.48\columnwidth}
\begin{lstlisting}[numbers=left]
var sum = 0.0
var index = 0
while(index < RSize) {
  var rec = null
  while(index < RSize) {
    val elem = R(index)
    index += 1
    if(elem.A < 10) {
      rec = elem
      break
    }
  }
  if(rec == null)
    break
  sum += rec.B
}
return sum
\end{lstlisting}
\subcaption{Inlined query in pull engine.}
\label{fig:pull-inlined-example}
\end{minipage}
\hspace{.03\columnwidth}
\begin{minipage}[c][\figcomplheightup][t]{.48\columnwidth}
\begin{lstlisting}
var sum = 0.0
var index = 0
while(index < RSize) {


  val rec = R(index)
  index += 1
  if(rec.A < 10) 
    
    
    
    
    
    
    sum += rec.B
}
return sum
\end{lstlisting}
\subcaption{Inlined query in push engine.}
\label{fig:push-inlined-example}
\end{minipage}
\begin{minipage}[c][\figcomplheightup][t]{.48\columnwidth}
\centering
\includegraphics[height=4.3cm]{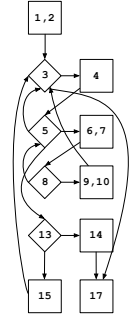}
\subcaption{The CFG of the inlined query in pull engine.}
\label{fig:pull-inlined-cfg}
\end{minipage}
\hspace{.03\columnwidth}
\begin{minipage}[c][\figcomplheightup][t]{.48\columnwidth}
\centering
\includegraphics[height=4.3cm]{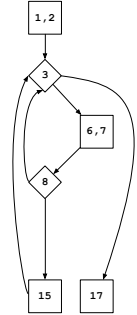}
\subcaption{The CFG of the inlined query in push engine.}
\label{fig:push-inlined-cfg}
\end{minipage}
\caption{Specialized version of the example query in pull and push engines and the corresponding control-flow graphs (CFG).}
\figurebottom
\label{fig:pushpull-inlined-example}
\end{figure}

With the advent of in-memory DBMSes, the code layout of the instructions becomes a very important factor. As a result, query compilation uses code generation and compilation techniques in order to inline virtual functions and further specialize the code to improve cache locality~\cite{ferry-2, DBLP:journals/pvldb/AhmadK09, DBLP:conf/pods/Koch10, krikellas, Neumann11, kochmanifesto, dbtoaster, legobase, DBLP:journals/debu/ViglasBN14, crotty2015tupleware, Nagel:2014:CGE:2732977.2732984, karpathiotakis2015just, spark-sql, dblablb, karpathiotakis2016fast}. As a result of that, the code pattern used in each pipelining algorithm really matters. Hence, it is important to investigate the performance of each pipelining algorithm for different workloads.

Figure~\ref{fig:pull-inlined-example} shows the inlined pull-engine code for the example SQL query given in Figure~\ref{fig:pushpull-example}. Note that for the selection operator, we need an additional \code{while} loop. 
This additional loop creates more branches in the generated code, which makes the control-flow graph (CFG) more complicated.
Figure~\ref{fig:pull-inlined-cfg} demonstrates the CFG of the inlined pull-engine code. Each rectangle in this figure corresponds to a block of statements, whereas diamonds represent conditionals. The edges between these nodes represent the execution flow. The backward edges represent the jumps inside a loop.
This complicated CFG makes the code harder to understand and optimize for the optimizing compiler. As a result, during the runtime execution, performance degrades mainly because of the worse branch prediction.

Figure~\ref{fig:push-inlined-example} shows the specialized query for a push engine of the previous example SQL query. The selection operator here is summarized in a single \code{if} statement. As a result, the CFG for the inlined push-engine code is simpler in comparison with the one for pull engine, as it is demonstrated in Figure~\ref{fig:push-inlined-cfg}. This makes the reasoning and optimization easier for the underlying optimizing compiler, leading to better branch prediction during runtime execution. 

Up to now, there is no separation of the concept of pipelining from the associated specializations. For example, HyPer~\cite{Neumann11} is in essence a push engine which uses compiler optimizations by default, without identifying the individual contributions to performance by these two factors. As another example, LegoBase~\cite{legobase} assumes that a push engine is followed by operator inlining, whereas the pull engine does not use operator inlining~\cite{legobase-errata}. On the other hand, there is no comparison between an inlined pull engine -- we suspect Hekaton~\cite{Diaconu:2013:HSS:2463676.2463710} to be of that class -- with a push-based inlined engine in the same environment. Hence, there is no comparison between pull and push engines which is under completely fair experimental conditions, sharing environment and code base to the maximum degree possible. In Section~\ref{sec:exp}, we attempt such a fair comparison.

Furthermore, na\"ively compiling the pull engine does not lead to good performance. This is because a na\"ive implementation of the iterator model does not take into account the number of \code{next} function calls. For example, the na\"ive implementation of the selection operator invokes the \code{next} method of its source operator twice, as it is demonstrated below:

\begin{lstlisting}[numbers=left]
class SelectOp[R] (p: R => Boolean) {
  def next(): R = {
    var elem: R = source.next()
    while(elem != null && !p(elem)) {
      elem = source.next()
    }
    elem
  }
}
\end{lstlisting}

The first invocation is happening before the loop for the initialization (line 3), and the second invocation is inside the loop (line 5). Inlining can cause an explosion of the code size, which can lead to worse instruction cache behavior. Hence, it is important to take into account these concerns while implementing query engines. For example, our implementation of the selection operator in a pull-based query engine invokes the \code{next} method of its source operator only once by changing the shape of the \code{while} loop (c.f. Figure~\ref{fig:pull-engine}). Section~\ref{sec:exp} shows the impact of this {\em \smartpull} implementation of pull engines.
\begin{figure*}[t]
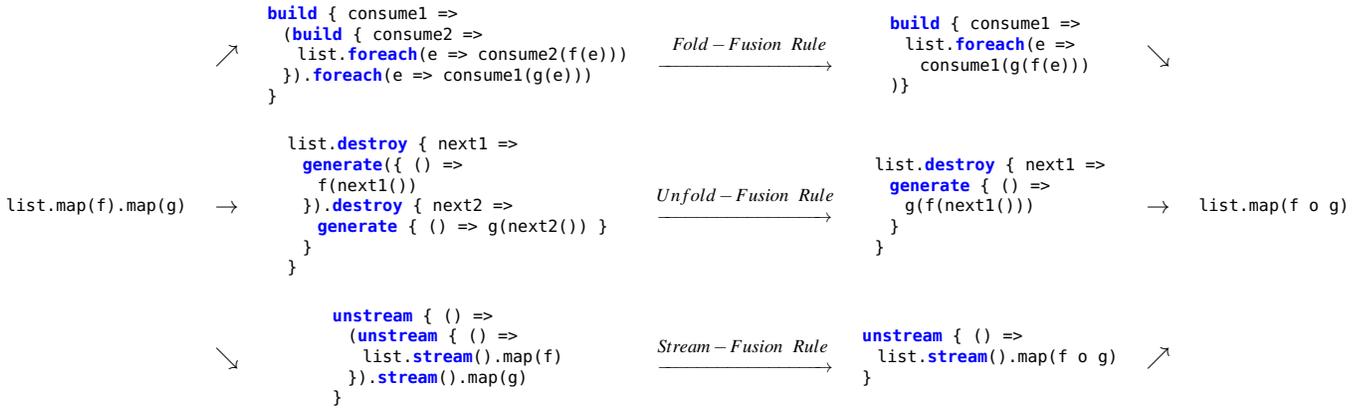

\def\arrowwidth{\hspace{2.2cm}}
\centering
\begin{tabular}{c c c c c c c}
&
$\nearrow$
&
\begin{lstlisting}
build { consume1 => 
  (build { consume2 => 
    list.foreach(e => consume2(f(e)))
  }).foreach(e => consume1(g(e)))
}
\end{lstlisting}
&
\begin{lstlisting}
$\hspace{0.2cm}Fold-Fusion\ Rule$
$\xrightarrow{\arrowwidth}$
\end{lstlisting}
&
\begin{lstlisting}
build { consume1 => 
  list.foreach(e => 
    consume1(g(f(e)))
)}
\end{lstlisting}
&
$\searrow$
&
\\
\\
\begin{lstlisting}
list.map(f).map(g)
\end{lstlisting}
&
$\rightarrow$
&
\begin{lstlisting}
list.destroy { next1 => 
  generate({ () => 
    f(next1()) 
  }).destroy { next2 => 
    generate { () => g(next2()) }
  }
}
\end{lstlisting}
&
\begin{lstlisting}
$Unfold-Fusion\ Rule$
$\xrightarrow{\arrowwidth}$
\end{lstlisting}
&
\begin{lstlisting}
list.destroy { next1 => 
  generate { () =>
    g(f(next1()))
  }
}
\end{lstlisting}
&
$\rightarrow$
&
\begin{lstlisting}
list.map(f o g)
\end{lstlisting}
\\
\\
&
$\searrow$
&
\begin{lstlisting}
unstream { () => 
  (unstream { () => 
    list.stream().map(f)
  }).stream().map(g)
}
\end{lstlisting}
&
\begin{lstlisting}
$Stream-Fusion\ Rule$
$\xrightarrow{\arrowwidth}$
\end{lstlisting}
&
\begin{lstlisting}
unstream { () => 
  list.stream().map(f o g)
}
\end{lstlisting}
&
$\nearrow$
&
\end{tabular}

\caption{Different fusion techniques on a simple example.}
\figurebottom
\label{fig:example-loop-fusion-simple}
\end{figure*}
\section{Loop Fusion in Collection Programming}
\label{sec:fusion}
Collection programming APIs are getting more and more popular. Ferry~\cite{ferry-1, ferry-2} and LINQ~\cite{linq} use such an API to seemlessly integrate applications with database back-ends. Spark \textit{RDDs}~\cite{rdd} use the same operations as collection programming APIs.
Also, functional collection programming abstractions exist in main-stream programming languages such as Scala, Haskell, and recently Java 8. The theoretical foundation of such APIs is based on Monad Calculus and Monoid Comprehensions~\cite{monad-calc-1, monad-calc-2, monad-comprehension, query-comprehension, query-comprehension-2, monoid-comprehension}.

Similar to query engines, the declarative nature of collection programming comes with a price. Each collection operation performs a computation on a collection and produces a transformed collection. A chain of these invocations results in creating unnecessary intermediate collections.

Loop fusion or Deforestation~\cite{deforestation} removes the intermediate collections in collection programs. This transformation is a nonlocal and brittle transformation which is difficult to apply to impure functional programs (i.e., in languages which include imperative features) and is thus absent from mainstream compilers for such languages. In order to provide a practical implementation, one can restrict the language to a pure functional DSL for which the fusion rules can be applied locally. These approaches are known as {\em short-cut}\/ deforestation, which remove intermediate collections using local transformations instead of global transformations. This makes it more realistic for them to be integrated into real compilers; short-cut approaches have been successfully implemented in the context of  Haskell~\cite{Svenningsson:2002:SFA:581478.581491, Coutts07streamfusion, foldr-fusion-1} and Scala-based DSLs~\cite{fold-based-fusion, dblablb}. 

Next, we present two approaches for short-cut deforestation in the order they were discovered.
Both approaches employ two methods of ``collection'' micro-instructions each, to which a large number of collection operations can be mapped. This allows to implement fusion using very few rewrite rules (in terms of these micro-instructions).

\subsection{Fold Fusion}
In this approach, every collection operation is implemented using two constructs: 1) the \code{build} method for \textit{producing} a collection, and 2) the \code{foldr} method for \textit{consuming} a collection. Some methods such as \code{map}, which transform a collection, use both of these constructs for consuming the given collection and producing a new collection. However, some methods such as \code{sum}, which produce an aggregated result from a collection, require only the \code{foldr} method for consuming the given collection.

We consider an imperative variant of this algorithm, in which the \code{foldr} method is substituted by \code{foreach}. The main difference is that the \code{foldr} method explicitly handles the state, whereas in the case of \code{foreach}, the state is handled internally and is not exposed to the interface.

Using Scala syntax, the signature of the \code{foreach} method on lists is as follows:

\vspace{0.2cm}
\begin{lstlisting}
class List[T] {
  def foreach(f: T => Unit): Unit
}
\end{lstlisting}

The \code{foreach} method consumes a collection by iterating over the elements of that collection and applying the given function to each element. The \code{build} function is the corresponding producer for the \code{foreach} method. This function produces a collection for which the \code{foreach} method applies the  \code{consumer} higher-order function to the function \code{f}. The signature of the \code{build} function is as follows:

\vspace{0.2cm}
\begin{lstlisting}
def build[T](consumer: (T => Unit) => Unit): List[T]
\end{lstlisting}

We illustrate the meanings of these two methods by an example. Consider the \code{map} method of a collection, which transforms a collection by applying a given function to each element. This method is expressed in the following way using the \code{build} and \code{foreach} functions:

\vspace{0.2cm}
\begin{lstlisting}
class List[T] {
  def map[S](f: T => S): List[S] = build { consume =>
    this.foreach(e => consume(f(e)))
  }
}
\end{lstlisting}

The implementation of several other collection operators using these two methods is given in Figure~\ref{fig:fold-fusion}.

After rewriting the collection operations using the \code{build} and \code{foreach} constructs, a pipeline of collection operators involves constructing intermediate collections. These intermediate collections can be removed using the following rewrite rule:

\fusionrule{Fold-Fusion Rule}{build(f1).foreach(f2)}{f1(f2)}

For example, there is a loop fusion rule for the \code{map} function, which fuses two consecutive \code{map} operations into one. More concretely, the expression \code{list.map(f).map(g)} is converted into \code{list.map(f o g)}. Figure~\ref{fig:example-loop-fusion-simple} demonstrates how the fold-fusion technique can derive this conversion by expressing the \code{map} operator in terms of \code{foreach} and \code{build}, following by application of the fold-fusion rule.

One of the key advantages of this approach is that instead of writing fusion rewrite rules for every combination of collection operations, it is sufficient to only express these operations in terms of the \code{build} and \code{foreach} methods. This way, instead of writing $O(n^2)$ rewrite rules for $n$ collection operations, it is sufficient to express these operations in terms of \code{build} and \code{foreach}, which is only $O(n)$ rewrite rules. Hence, this approach greatly simplifies the maintenance of the underlying compiler transformations~\cite{dblablb}.  

This approach successfully deforests most collection operators very well. However, it is not successful in the case of \code{zip} and \code{take} operations. The \code{zip} method involves iterating over two collections, which cannot be expressed using the \code{foreach} construct which iterates only over one collection. Hence, this approach  can deforest only one of the collections and for the other one, an intermediate collection must be created. Also, for the \code{take} method, there is no way to stop the iteration of the \code{foreach} method halfway to finish. Hence, the fold fusion technique does not perform well in these two cases. The next fusion technique solves the problem with these two methods.

\subsection{Unfold Fusion}
This is considered a dual approach to fold fusion.
Every collection operation is expressed in terms of the two constructs \code{generate}\footnote{We are presenting an imperative version of unfold fusion here; the purely functional version employs an \code{unfold} function instead of \code{generate}, and the approach derives its name from that.} and \code{destroy}, which have the following prototypes:



\vspace{0.2cm}
\begin{lstlisting}
class List[T] {
  def destroy[S](f: (() => T) => S): S
}
def generate[T](next: () => T): List[T]
\end{lstlisting}

The \code{destroy} method consumes the given list. Each element of this collection is accessible by invoking the \code{next} function available by the \code{destroy} method. 
The \code{generate} function generates a collection, that its elements are specified by the input function passed to this method. In the case of \code{map} operator, the elements of the result collection are the elements of the input collection after the \code{f} function being applied to them. 

The \code{map} method of collections is expressed in the following way using the \code{generate} and \code{destroy} methods:

\vspace{0.2cm}
\begin{lstlisting}
class List[T] {
  def map[S](f: T => S): List[S] = this.destroy { next =>
    generate { () =>
      val elem = next()
      if(elem == null) null
      else f(elem)
    }
  }
}
\end{lstlisting}

The implementation of some other collection operators using these two methods is given in Figure~\ref{fig:unfold-fusion}.

In order to remove the intermediate collections, the chain of intermediate \code{generate} and \code{destroy} can be removed. This fact is shown in the following transformation rule:

\fusionrule{Unfold-Fusion Rule}{generate(f1).destroy(f2)}{f2(f1)}

Figure~\ref{fig:example-loop-fusion-simple} demonstrates how this rule fuses the previous example, \code{list.map(f).map(g)} into \code{list.map(f o g)}. Note that the null checking statements, which are for checking the end of a list, are removed for brevity.

This approach introduces a recursive iteration for the \code{filter} operation. In practice, this can cause performance issues, however from a theoretical point of view the deforestation is applied successfully~\cite{Hinze:2010:TPF:2050135.2050137}. Also, this approach does not fuse operations on nested collections, which is beyond the scope of this paper.

\begin{table*}
\small
\begin{center}
  \begin{tabular}{| l | c | c |  c | c | c | c | c | c |  }
    \hline
    {\bf Operator Category} & {\bf Producer} & \multicolumn{6}{c|}{{\bf Transformer}} & {\bf Consumer} \\ \hline
    Query Operator &  Scan & Selection & Projection & OrderBy & Limit & Join* & Merge Join$^{\dagger}$  & Agg$^{\ddagger}$ \\ \hline
    Collection Operator & \code{List.fromArray} & \code{filter} & \code{map} & \code{sortBy} & \code{take} & \code{flatMap}* & \code{zip}$^{\dagger}$ &  \code{fold}$^{\ddagger}$ \\
    \hline
  \end{tabular}
  
  \end{center}
{\footnotesize * Nested loop join can be expressed using two nested flatMaps, but there is no equivalent for hash-based joins. Also, flatMaps can express nested collections, whereas in relational query engines every relation is considered to be flat.} \\
{\footnotesize$\dagger$ Merge join and \code{zip} have similar behavior, but their semantic is different.} \\
{\footnotesize$\ddagger$ An Agg operator representing a Group By is a transformer, whereas the one which folds into only a single result is a consumer.}
  
\caption{Mapping between query operators and collection operators}
\label{table:qop-colmethod}
\end{table*}
\begin{table}[t]
\small
\begin{center}
  \begin{tabular}{| l | l | l | }
    \hline
    {\bf Pipelined} & {\bf Object-Oriented} & {\bf Collection} \\
    {\bf Query Engines} & {\bf Design Pattern} & {\bf Loop Fusion} \\ \hline
    Pull Engine &  Iterator & Unfold fusion~\cite{Svenningsson:2002:SFA:581478.581491} \\ & & Stream fusion~\cite{Coutts07streamfusion} \\ \hline
    Push Engine & Visitor & Fold fusion~\cite{foldr-fusion-1} \\
    \hline
  \end{tabular}
  
  \end{center}
  
  \vspace{-2mm}
  
  \caption{Correspondence among pipelined query engines, object-oriented design patterns, and collection programming loop fusion.}
\label{table:pipeling-loopfusion}


\end{table}
\begin{figure*}[t!]
\def\figcomplheightup{11.2cm}
\def\figcomplheightdown{11.7cm}

\begin{minipage}[c][\figcomplheightup][t]{.29\textwidth}

\begin{lstlisting}

class ProjectOp[R, P](f: R => P) {
  def consume(e: R): Unit = 
    (*@ \MyColorBox[col2]{dest.consume(f(e))} @*)
}
class SelectOp[R](p: R => Boolean) {
  def consume(e: R): Unit =
    if(p(e))
      (*@ \MyColorBox[col1]{  dest.consume(e)} @*)
}
class AggOp[R, S](f: (R, S) => S) {
  var result = zero[S]
  def consume(e: R): Unit = {
    (*@ \MyColorBox[col2]{result = f(e, result)} @*)
  }
  def getResult: S = result
}
class HashJoinOp[R, R2]
  (leftHash: R => Int)
  (rightHash: R2 => Int)
  (cond: (R, R2) => Boolean) {
  val hm = new MultiMap[Int, R]()
  def consumeLeft(e: R): Unit = {
    (*@ \MyColorBox[col1]{hm.addBinding(leftHash(e) -> e)} @*)
  }
  def consumeRight(e: R2): Unit = {
    hm.get(rightHash(e)) match {
      case Some(list) =>
        for(l <- list) {
          if(cond(l, e)) {
            (*@ \MyColorBox[col2]{dest.consume(l.concat(e))} @*)
          }
        }
      case None => 
} } }
\end{lstlisting}
\subcaption{Push-based query engine}
\label{fig:push-engine}
\end{minipage}
\begin{minipage}[c][\figcomplheightup][t]{.39\textwidth}
\begin{lstlisting}
class QueryMonad[R] {
  def map[S](f: R => S) = build { consume =>
    for(e <- this) 
      (*@ \MyColorBox[col2]{consume(f(e))}  @*)
  }
  def filter(p: R => Boolean) = build { consume =>
    for(e <- this) 
      if(p(e))
      (*@ \MyColorBox[col1]{  consume(e)} @*)
  }
  def fold[S](zero: S)(f: (R, S) => S): S = {
    var result = zero
    for(e <- this) {
    (*@ \MyColorBox[col2]{result = f(e, result)} @*)
    }
    result
  }
  def hashJoin[R2](rightList: QueryMonad[R2])
    (leftHash: R => Int)
    (rightHash: R2 => Int)
    (cond: (R, R2) => Boolean) = build { consume =>
    val hm = new MultiMap[Int, R1]()
    for(e <- this) {
      (*@ \MyColorBox[col1]{hm.addBinding(leftHash(e) -> e)} @*)
    }
    for(e <- rightList) {
      hm.get(rightHash(e)) match {
        case Some(list) =>
          for(l <- list) {
            if(cond(l, e)) {
              (*@ \MyColorBox[col2]{consume(l.concat(e))} @*)
            }
          }
        case None => 
} } } }
\end{lstlisting}
\subcaption{Fold fusion of collections.}
\label{fig:fold-fusion}
\end{minipage}
\begin{minipage}[c][\figcomplheightup][t]{.32\textwidth}
\begin{lstlisting}

type Cont[T] = (T => Unit) => Unit

class Consumer[T]
  (val cont: Cont[T])
  extends QueryMonad[T] {
  def foreach(f: T => Unit): Unit = 
    cont(f)
}

def build[T](cont: Cont[T])
  : QueryMonad[T] = 
  new Consumer[T](cont)
\end{lstlisting}
\subcaption{The constructs for fold fusion.}
\label{fig:fold-fusion-constructs}
\vspace{2.3cm}
\begin{lstlisting}
build(f1).foreach(f2)

(*@ $\Big\downarrow$ (inline build definition) @*)

new Consumer(f1).foreach(f2)

(*@ $\Big\downarrow$ (inline foreach definition) @*)

f1(f2)
\end{lstlisting}
\subcaption{The derivation of the fold-fusion rule.}
\label{fig:fold-fusion-rule}
\end{minipage}
\\
\begin{minipage}[c][\figcomplheightdown][t]{.29\textwidth}
\begin{lstlisting}

class SelectOp[R](p: R => Boolean) {
  def next(): R = {
    var elem: R = null
    do {
      (*@ \MyColorBox[col1]{elem = source.next()} @*)
    } while (elem != null && !p(elem))
    elem
  }
}
class ProjectOp[R, P](f: R => P) {
  def next(): P = {
   (*@\MyColorBox[col2]{val elem = source.next()} @*)
    if(elem == null) null
    else f(elem)
  }
}
class AggOp[R, S]
  (f: (R, S) => S) {
  def next(): S = {
    var result = zero[S]
   (*@\MyColorBox[col1]{var elem:R = source.next()} @*)
    while(elem != null){
      (*@\MyColorBox[col2]{result = f(elem, result)} @*)
      (*@\MyColorBox[col1]{elem = source.next()} @*)
    }
    result
  }
}
class LimitOp[R](n: Int) {
  var count = 0
  def next(): R = {
    if(count < n) {
      count += 1
     (*@\MyColorBox[col2]{source.next()} @*)
    } else {
      null
} } }
\end{lstlisting}
\subcaption{Pull-based query engine}
\label{fig:pull-engine}
\end{minipage}
\begin{minipage}[c][\figcomplheightdown][t]{.39\textwidth}
\begin{lstlisting}
class QueryMonad[R] {
  def filter(p: R => Boolean) = destroy { next =>
    generate { () =>
      var elem: R = null
      do {
        (*@\MyColorBox[col1]{elem = next()} @*)
      } while(elem != null && !p(elem))
      elem
    }
  }
  def map[P](p: R => P) = destroy { next =>
    generate { () =>
     (*@\MyColorBox[col2]{val elem = next()} @*)
      if(elem == null) null
      else f(elem)
    }
  }
  def fold[S](zero: S)
    (f: (R, S) => S): S = 
    destroy { next =>
      var result = zero
     (*@\MyColorBox[col1]{var elem:R = next()} @*)
      while(elem != null){
        (*@\MyColorBox[col2]{result = f(elem, result)} @*)
        (*@\MyColorBox[col1]{elem = next()} @*)
      }
      result
    }

  def take(n: Int) = {
    var count = 0
    destroy { next =>
      if(count < limit) {
        count += 1
       (*@\MyColorBox[col2]{next()} @*)
      } else {
        null
} } } }
\end{lstlisting}
\subcaption{Unfold fusion of collections.}
\label{fig:unfold-fusion}
\end{minipage}
\begin{minipage}[c][\figcomplheightup][t]{.32\textwidth}
\begin{lstlisting}

type Gen[T] = () => T

type Dest[T, S] = (Gen[T] => S) => S

class DestroyGen[T]
  (val next: Gen[T])
  extends QueryMonad[T] {
  def destroy[S](f: Gen[T] => S): S = 
    f(next)
}

def generate[T](next: Gen[T])
  : QueryMonad[T] = 
  new DestroyGen[T](next)
\end{lstlisting}
\subcaption{The constructs for unfold fusion.}
\label{fig:unfold-fusion-constructs}
\vspace{2.3cm}
\begin{lstlisting}
generate(f1).destroy(f2)

(*@ $\Big\downarrow$ (inline generate definition) @*)

new DestroyGen(f1).destroy(f2)

(*@ $\Big\downarrow$ (inline destroy definition) @*)

f2(f1)
\end{lstlisting}
\subcaption{The derivation of the unfold-fusion rule.}
\label{fig:unfold-fusion-rule}
\end{minipage}
\caption{Comparison of pull-based and push-based pipelining and loop fusion algorithms; code snippets in Scala.}
\figurebottom
\label{fig:pipelining-fusion}
\end{figure*}
\subsection{Loop Fusion is Operator Pipelining}
\label{sec:cor}

By chaining query operators, one can express a given (say, SQL) query. Similarly, a given collection program can be expressed using a pipeline of collection operators. The relationship between relational queries and collection programs has been well studied. In particular,
one can establish a precise correspondence between relational query plans and a class of collection programs \cite{PG88}.

Operators can be divided into three categories: 1) The operators responsible for \textit{producing} a collection from a given source (e.g., a file or an array), 2) The operators which \textit{transform} the given collection to another collection, and 3) The \textit{consumer} operators which aggregate the given collection into a single result. 

The mapping between query operators and collection operators is summarized in Table~\ref{table:qop-colmethod}. Most join operators do not have a directly corresponding collection operator with two exceptions: Nested loop joins can be expressed using nested \code{flatMap}s and the \code{zip} collection operator is very similar to the merge join query operator. Both operators need to traverse two input sequences in parallel. For the rest of join operators, we extend collection programming with join operators (e.g. \code{hashJoin}, \code{semiHashJoin}, etc.). A similar mapping between the LINQ~\cite{linq} operators and Haskell lists is shown in Steno~\cite{Murray:2011:SAO:1993498.1993513}. Note that we do not consider nested collections here, although straightforward to support in collection programming, in order to emphasize similarity with relational query engines.

Pipelining in query engines is analogous to loop fusion in collection programming. Both concepts remove the intermediate relations and collections, which break the stream pipeline. Also, pipelining in query engines matches well-known design patterns in object-oriented programming~\cite{vlissides1995design}. The correspondence among pipelining in query engines, design patterns in object-oriented languages, and loop fusion in collection programming is summarized in Table~\ref{table:pipeling-loopfusion}.

\para{Push Engine = Fold Fusion} There is a similarity between the Visitor pattern and fold fusion. On one hand it has been proven that the Visitor design pattern corresponds to the Church-encoding~\cite{bohm1985automatic} of data types~\cite{Buchlovsky2006309}. On the other hand, the \code{foldr} function on a list corresponds to the Church-encoding of lists in $\lambda$-calculus~\cite{pierce2002types, coroutine-fusion}. Hence, both approaches eliminate intermediate results by converting the underlying data structure into its Church-encoding. In the former case, specialization consists of inlining, which results in removing (virtual) function calls. In the latter case, the fold-fusion rule and $\beta$-reduction are performed to remove the materialization points and inline the $\lambda$ expressions. The correspondence between these two approaches is shown in Figure~\ref{fig:pipelining-fusion} (compare (a) vs.\ (b)). The invocations of the \code{consume} method of the destination operators in the push engine corresponds to the invocation of the \code{consume} function which is passed to the \code{build} operator in fold fusion.

\para{Pull Engine = Unfold Fusion} In a similar sense, the Iterator pattern is similar to unfold fusion. Although the category-theoretic essence of the iterator model was studied before~\cite{gibbons2009essence}, there is no literature on the direct correspondence between the \code{unfold} function and the Iterator pattern. However, Figure~\ref{fig:pipelining-fusion} shows how a pull engine is similar to unfold fusion (compare Figure~\ref{fig:pipelining-fusion} (e) vs. (f)), to the best of our knowledge for the first time. Note the correspondence between the invocation of the \code{next} function of the source operator in pull engines, and the invocation of the \code{next} function which is passed to the \code{destroy} operator in unfold fusion, which is highlighted in the figure.

\section{An Improved Pull-Based Engine}
\label{sec:stream}
In this section, we first present yet another loop-fusion technique for collection programs. Then, we suggest a new pull-based query engine inspired by this fusion technique based on the correspondence between queries and collection programming.

\subsection{Stream Fusion}
In functional languages, loops are expressed as recursive functions. Reasoning about recursive functions is very hard for optimizing compilers.
Stream fusion tries to solve this issue by converting all recursive collection operations to non-recursive stream operations. To do so, first all collections are converted to streams using the \code{stream} method. Then, the corresponding method on the stream is invoked which results in a transformed stream. Finally, the transformed stream is converted back to a collection by invoking the \code{unstream} method.

The signature of the \code{unstream} and \code{stream} methods is as follows:
\begin{lstlisting}
def unstream[T](next: () => Step[T]): List[T]
class List[T] {
  def stream(): Step[T]
}
\end{lstlisting}

For example, the \code{map} method is expressed in using these two methods as:
\begin{lstlisting}
class List[T] {
  def map[S](f: T => S): List[S] = unstream { () =>
    this.stream().map(f)
  }
}
\end{lstlisting}
The \code{stream} method converts the input collection to an intermediate stream, which is specified by the \code{Step} data type. The function \code{f} is applied to this intermediate stream using the \code{map} function of the \code{Step} data type. Afterwards, the result stream is converted back to a collection by the \code{unstream} method.

As discussed before, one of the main advantages of the intermediate stream, the \code{Step} data structure, is that its operations are mainly non-recursive. This simplifies the task of the optimizing compiler to further specialize the program. The implementation of several methods of the \code{Step} data structure is given in Figure~\ref{fig:stream-engine-step}.

\begin{figure*}[t!]
\def\figcomplheight{11cm}

\begin{minipage}[c][\figcomplheight][t]{.30\textwidth}
\begin{lstlisting}

class SelectOp[R](p: R => Boolean) {
  def stream(): Step[R] = {
   (*@ \MyColorBox[col2]{source.stream().filter(p)} @*)
  }
}
class ProjectOp[R, P](f: R => P) {
  def stream(): Step[P] = {
   (*@ \MyColorBox[col1]{source.stream().map(f)} @*)
  }
}
class AggOp[R, S](f: (R, S) => S) {
  def stream(): Step[S] = {
    var result = zero[S]
    var done = false
    while(!done){
      (*@ \MyColorBox[col2]{source.stream().fold(}@*)
         e => { result = f(e, result) },
         () => ,
         () => { done = true }
       )
    }
    return result
  }
}
class LimitOp[R](n: Int) {
  var count = 0
  def stream(): Step[R] = {
    if(count < n) {
     (*@\MyColorBox[col1]{source.stream().map(e => \{ }@*)
        count += 1 
        e
      })
    } else {
      Done
} } }
\end{lstlisting}
\subcaption{Stream-Fusion Query Engine}
\label{fig:loop-fusion-example-complicated-mchl}
\end{minipage}
\begin{minipage}[c][\figcomplheight][t]{.35\textwidth}
\begin{lstlisting}
class QueryMonad[R] {
  def filter(p: R => Boolean) = {
    unstream { () =>
      (*@\MyColorBox[col2]{stream().filter(p)}@*)
    }
  }
  def map[P](f: R => P) = {
    unstream { () =>
      (*@\MyColorBox[col1]{stream().map(f)}@*)
    }
  }
  def fold[S](z: S)(f: (R, S) => S): S = {
    unstream { () =>
      var result = zero[S]
      var done = false
      while(!done){
       (*@\MyColorBox[col2]{stream().fold(}@*)
          e => { result = f(e, result) },
          () => ,
          () => { done = true }
        )
      }
      return result
    }
  }
  def take(n: Int) = {
    var count = 0
    unstream { () =>
      if(count < n) {
       (*@\MyColorBox[col1]{stream().map(e => \{ }@*)
          count += 1 
          e
        })
      } else {
        Done
} } } }

\end{lstlisting}
\subcaption{Stream fusion of collections.}
\label{fig:unfold-fusion2}
\end{minipage}
\begin{minipage}[c][\figcomplheight][t]{.35\textwidth}
\begin{lstlisting}

trait Step[T] {
  def filter(p: T => Boolean): Step[T]
  def map[S](f: T => S): Step[S]
  def fold[S](yld: T => S,
    skip: () => S, done: () => S): S
}

case class Yield[T](e: T) extends Step[T] {
  def filter(p: T => Boolean) = 
    if(p(e)) Yield(e) else Skip
  def map[S](f: T => S) = 
    Yield(f(e))
  def fold[S](yld: T => S, 
    skip: () => S, done: () => S): S =
    yld(e)
}

case object Skip extends Step[Nothing] {
  def filter(p: Nothing => Boolean) = 
    Skip
  def map[S](f: Nothing => S) = 
    Skip
  def fold[S](yld: Nothing => S, 
    skip: () => S, done: () => S): S =
    skip()
}

case object Done extends Step[Nothing] {
  def filter(p: Nothing => Boolean) = 
    Done
  def map[S](f: Nothing => S) = 
    Done
  def fold[S](yld: Nothing => S,
    skip: () => S, done: () => S): S =
    done()
}

\end{lstlisting}
\subcaption{The operations of the Step data type.}
\label{fig:stream-engine-step}
\end{minipage}
\caption{Stream-based query engine and the stream fusion technique.}
\figurebottom
\label{fig:pipelining-fusion2}
\end{figure*}

Such transformations do not result in direct performance gain -- they may even degrade performance. This is because of the intermediate conversions between streams and collections. However, these intermediate conversions can be removed using the following rewrite rule:

\fusionrule{Stream-Fusion Rule}{unstream(() => e).stream()}{e}

Figure~\ref{fig:example-loop-fusion-simple} demonstrates how the stream fusion technique transforms \code{list.map(f).map(g)} into \code{list.map(f o g)}. Note that for the \code{Step} data type, the \code{step.map(f).map(g)} expression is equivalent to \code{step.map(f o g)}.

The idea behind stream fusion is very similar to unfold fusion. The main difference is the \code{filter} operator. Stream fusion uses a specific value, called \code{Skip},  to implement the \code{filter} operator. This is in contrast with the unfold fusion approach for which the \code{filter} operator is implemented using an additional nested \code{while} loop for skipping the unnecessary elements. Hence, stream fusion solves the practical problem of unfold fusion associated with the \code{filter} operator.

Next, we define a new pipelined query engine based on the ideas of stream fusion.

\subsection{Stream-Fusion Engine}
The proposed query engine follows the same design as the iterator model. Hence, this engine is also a pull engine. However, instead of invoking the \code{next} method, this engine invokes the \code{stream} method, which returns a wrapper object of type \code{Step}. We refer to our proposed engine as the \textit{stream-fusion engine}.

As we mentioned in Section~\ref{sec:engine-pull}, one of the main practical problems with a pull engine is the case of the selection operator. In this case, an operator waits until the selection operator returns the next satisfying element. The proposed engine solves this issue by using the \code{Skip} object which specifies that the current element should be ignored. Hence, selection operators are no longer a blocker for their destination operator.

The correspondence between the stream fusion algorithm and the stream-fusion engine is shown in Figure~\ref{fig:pipelining-fusion2}. Every query operator provides an appropriate implementation for the \code{stream} method, which invokes the \code{stream} method of the source operator to request the next element. Similarly, stream fusion uses the \code{stream} method to fetch the next element. Then, by invoking the \code{unstream} method the generated stream is converted back to a collection.

From a different point of view, a push engine can be expressed using a \code{while} loop and a construct for skipping to the next iteration (e.g. \code{continue}). By nature, it is impossible for a push-based engine to finish the iteration before the producer's \code{while} loop finishes its job. 
In contrast, a pull engine is generally expressible using a \code{while} loop and a construct for terminating the execution of the \code{while} loop (e.g. \code{break}). This is because of the demand-driven nature of pull engines. However, in a pull-based engine there is no way to skip an iteration. As a result, this should be expressed using a nested \code{while} loop which results in performance issues (c.f. Section~\ref{sec:engine}). 

The stream-fusion engine solves the mentioned problem by adding a \code{Skip} construct which results in skipping to the next iteration. This has an equivalent effect to the \code{continue} construct. Table~\ref{table:engine-while} summarizes the differences among the aforementioned query engines.

Consider a relation of two elements for which we select its first element and the second element is filtered out. The first call to the \code{stream} method of the selection operator in the stream-fusion engine, produces a \code{Yield} element, which contains the first element of the relation. The second invocation of the same method returns a \code{Skip} element, specifying that this element, which is the second element of the relation, is filtered out and should be ignored. The next invocation of this method, results in a \code{Done} element, denoting that there is no more element to be produced by the selection operator. The \code{Done} value has the same role as the \code{null} value in the pull engine.

\begin{figure}[t]
\def\figcomplheightup{5.9cm}
\begin{minipage}[c][\figcomplheightup][t]{.48\columnwidth}
\begin{lstlisting}
var index = 0
var sum = 0.0
while(true) {
  val step1 = 
    if(index < RSize) {
      val rec = R(index)
      index += 1
      Yield(rec)
    } else 
      Done
  step1.filter(x => x.A < 10)
  .map(x => x.B)
  .fold(x => sum += x,
         () => ,
         () => break)
}
return sum
\end{lstlisting}
\subcaption{Inlined query in stream-fusion engine without further specializations.}
\label{fig:stream-inlined-example-1}
\end{minipage}
\begin{minipage}[c][\figcomplheightup][t]{.50\columnwidth}
\begin{lstlisting}
var index = 0
var sum = 0.0
while(true) {

  if(index < RSize) {
    val rec = R(index)
    index += 1
    
    
    
    if(rec.A < 10) 
      sum += rec.B
  } 
  else 
    break
}
return sum
\end{lstlisting}
\subcaption{Inlined query in stream-fusion engine by inlining the visitor model of \code{Step}.}
\label{fig:stream-inlined-example-2}
\end{minipage}
\caption{Specialization of the example query in stream-fusion engine.}
\figurebottom
\label{fig:stream-inlined-example}
\end{figure}

The specialized version of the example query (which was introduced in Figure~\ref{fig:pushpull-example}) based on the stream-fusion engine is shown in Figure~\ref{fig:stream-inlined-example-1}. The code is as compact as the push engine code. However, it suffers from some performance problems due to the intermediate \code{Step} objects created. The next section discusses implementation aspects and the optimizations needed for tuning the performance of the stream-fusion engine.

\begin{table}[t]
\small
\begin{center}
  \begin{tabular}{| l | l | }
    \hline
    {\bf Pipelined Query Engines} & {\bf Looping Constructs} \\ \hline
    Push Engine & \code{while} + \code{continue} \\ \hline
    Pull Engine & \code{while} + \code{break} \\ \hline
    Stream-Fusion Engine & \code{while} + \code{break} + \code{continue} \\ \hline
  \end{tabular}
  
  \end{center}
  
  \caption{The supported looping constructs by each pipelined query engine.}
\label{table:engine-while}

  \figurebottom

\end{table}
\section{Implementation}
\label{sec:impl}
In this section, we discuss the implementation of the presented query engines. First, we discuss how the fusion rules are implemented for each approach. Then, we show how the problem associated with intermediate objects is resolved for the stream-fusion engine.

We used the open-source DBLAB framework~\cite{dblablb}~\footnote{http://github.com/epfldata/dblab} to implement different query engines and the associated optimizations. This framework allows us to implement these engines in the high-level programming language Scala. The input programs can either be expressed using physical (relational algebra-style) query plans in the \qopl{} language or collection programming using the \qml{} language. Furthermore, we implement the optimizations using rewrite rules which are provided by the transformation framework of DBLAB.

We implemented the collection programming operations and the corresponding loop fusion techniques. Due to the equivalence which was shown in Section~\ref{sec:cor} between query engines and collection programming, it is clear how they can be implemented for query engines. As a result, the experimental results presented in the next section for different fusion techniques matches the results for different approaches for pipelined query engines. Next, we discuss how the fusion rules for different loop fusion algorithms can be expressed in this framework.

\subsection{Fusion By Inlining}
As mentioned in Section~\ref{sec:cor}, in loop fusion techniques, the fusion rule is expressed as a local transformation rule which is applied as an extension to the host language compiler (which is GHC~\cite{jones1993glasgow} in the case of the mentioned papers). In this section, we show how these fusion rules are implemented by only using inlining. This was proposed for implementing fold fusion in Scala~\cite{fold-based-fusion}. Here, we use a similar approach for other fusion techniques.

Figure~\ref{fig:fold-fusion-constructs} shows the definition of the \code{build} operator. By inlining the definition of this operator, an object of type \code{QueryMonad} is created. The \code{foreach} method of this object applies the higher-order function passed to the \code{build} method (\code{f1}) to the input parameter of the \code{foreach} method (\code{f2}). By inlining this \code{foreach} method, we derive the same rule as the fold-fusion rule which was introduced in Section~\ref{sec:fusion}. This derivation is shown in Figure~\ref{fig:fold-fusion-constructs}. The constructs and derivation of unfold fusion are shown in Figure~\ref{fig:unfold-fusion-constructs} and Figure~\ref{fig:unfold-fusion-rule}. Stream fusion follows a similar pattern which is given in Figure~\ref{fig:stream-fusion-constructs} and Figure~\ref{fig:stream-fusion-rule}. 

\begin{figure}[t]
\begin{lstlisting}

type GenStream[T] = () => Step[T]

class Streamer[T]
  (val next: GenStream[T])
  extends QueryMonad[T] {
  def stream(): Step[T] = 
    next()
}

def unstream[T](next: GenStream[T])
  : QueryMonad[T] = 
  new Streamer[T](next)
\end{lstlisting}
\caption{The constructs for stream fusion.}
\label{fig:stream-fusion-constructs}
\vspace{0.5cm}
\begin{lstlisting}
unstream(() => e).stream()

(*@ $\Big\downarrow$ (inline unstream definition) @*)

new Streamer(() => e).stream()

(*@ $\Big\downarrow$ (inline stream definition) @*)

e
\end{lstlisting}
\caption{The derivation of the stream-fusion rule.}
\label{fig:stream-fusion-rule}
\end{figure}

Next, we discuss the problematic creation of intermediate objects by the stream-fusion engine, as well as our solution.

\subsection{Removing Intermediate Results}
Although the stream-fusion engine removes intermediate relations, it creates intermediate \code{Step} objects. There are two problems with these intermediate objects. First, the \code{Step} data type operations are virtual calls. This causes poor cache locality and degrades the performance. Second, normally these intermediate objects lead to heap allocations. This causes more memory consumption and a worse runtime. This is why the original stream fusion approach is dependent on optimizations provided by its source language compiler (i.e., the GHC~\cite{jones1993glasgow} compiler). Implementing an effective version of it for other languages requires supporting similar optimizations supported by the GHC compiler.

The first problem with virtual calls can be solved by rewriting the \code{Step} operations by enumerating all cases for the \code{Step} object. This is possible because there are only three possible concrete cases (1. \code{Yield} 2. \code{Skip} 3. \code{Done}) for this data type. To do so, one can use if-statements. In functional languages, the pattern matching feature can be used. Although this approach solves the first problem, still there are heap allocations which are not removed.

The good news is that these heap allocations can be converted to stack allocations. This is because the created objects are not escaping their usage scope. For example, these objects are not copied into an array and not used as an argument to a function. This fact can be verified by the well-known compilation technique of \textit{escape analysis}~\cite{choi1999escape}. Based on that, the heap allocations can be converted to stack allocations.

\begin{figure}[t!]
\begin{lstlisting}
trait StepVisitor[T] {
  def yld(e: T): Unit
  def skip(): Unit
  def done(): Unit
}

trait Step[T] { self =>
  def __match(v: StepVisitor[T]): Unit
  def filter(p: T => Boolean): Step[T] =
    new Step[T] {
      def __match(v: StepVisitor[T]): Unit =
        self.__match(new StepVisitor[T] {
          def yld(e: T): Unit =
            if (p(e)) v.yld(e) else v.skip()
          def skip(): Unit = v.skip()
          def done(): Unit = v.done()
        })
    }
  def map[S](f: T => S): Step[S] =
    new Step[S] {
      def __match(v: StepVisitor[S]): Unit =
        self.__match(new StepVisitor[T] {
          def yld(e: T): Unit = v.yld(f(e))
          def skip(): Unit = v.skip()
          def done(): Unit = v.done()
        })
    }
}

case class Yield[T](e: T) extends Step[T] {
  def __match(v: StepVisitor[T]): Unit = v.yld(e)
}
case object Skip extends Step[Nothing] {
  def __match(v: StepVisitor[Nothing]): Unit = v.skip()
}
case object Done extends Step[Nothing] {
  def __match(v: StepVisitor[Nothing]): Unit = v.done()
}
\end{lstlisting}
\caption{Step data type implemented using the Visitor pattern.}
\label{fig:stream-engine-step-visitor}
\end{figure}

The compiler optimizations can go further and remove the stack allocations as well. Instead of the stack allocation for creating a \code{Step} object, the fields necessary to encode this type are converted to local variables. Hence the \code{Step} abstraction is completely removed. This optimization is known as \textit{scalar replacement} in compilers.

From a different point of view, removing the intermediate \code{Step} objects is a similar problem to removing the intermediate relations and collections in query engines and collection programming. Hence, one can borrow similar ideas and apply it for the \code{Step} objects in a fine-grained granularity. 

To do so, we implement a variant of the \code{Step} data type using the Visitor pattern. As we discussed in Section~\ref{sec:cor}, this is similar to the Church-encoding of data types. This encoding results in \textit{pushing} \code{Step} objects down the pipeline. Hence, the stream-fusion engine implements a pull engine on a coarse-grained level and pushes the tuples on a fine-grained level. The Visitor pattern version of the \code{Step} data type is shown in Figure~\ref{fig:stream-engine-step-visitor}.

The result of applying this enhancement to our working example is shown in Figure~\ref{fig:stream-inlined-example-2}. By comparing this code to the code produced by a push engine, we see a clear similarity. First, there are no more additional virtual calls associated with the \code{Step} operators. Second, there is no more materialization of the intermediate \code{Step} objects. Finally, similar to push engines, the produced code does not contain any additional nested \code{while} loop for selection, hence it is easier to understand and optimize by an underlying compiler.

As an alternative implementation, one can implement the \code{Step} data type as a \textit{sum} type, a type with different distinct cases in which an object can be one and only one of those cases. Hence, the implementation of the \code{Step} methods can use the pattern matching feature of the Scala programming language. However, it has been proven that the Visitor pattern is a way to encode the sum types in object-oriented languages~\cite{Buchlovsky2006309}.
On the other hand, pattern matching in Scala is a way to express the Visitor pattern~\cite{Emir:2007:MOP:2394758.2394779}. Hence, from a conceptual point of view there is no difference between these implementations~\cite{ Hofer:2010:MDL:1868294.1868307}.

\section{Experimental Results}
\label{sec:exp}
For the experimental evaluation, we use a server-type x86 machine equipped with
two Intel Xeon E5-2620 v2 CPUs running at 2GHz each, 256GB of DDR3 RAM at
1600Mhz and two commodity HDDs of 2TB.
The operating system is Red Hat Enterprise 6.7.

\begin{figure}[t]
\centering
\includegraphics[width=\columnwidth]{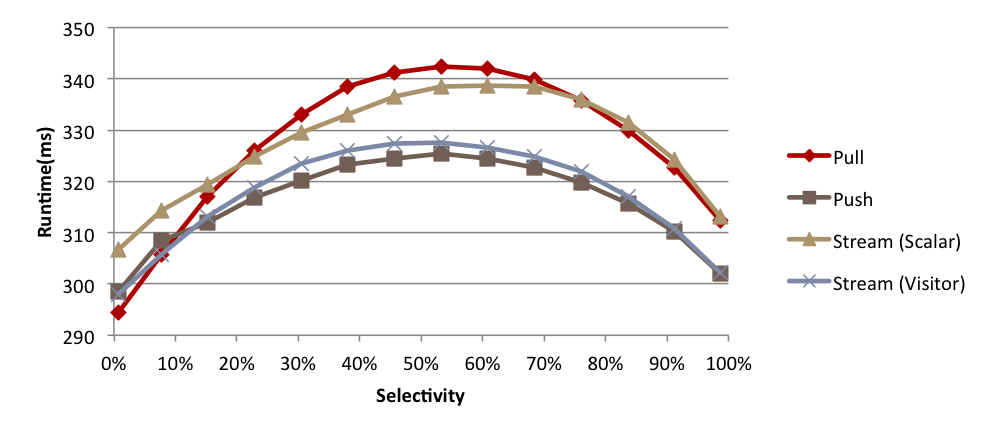}
\caption{Sensitivity of query engines to selectivity.}
\figurebottom
\label{fig:selectivity}
\end{figure}

Our query compiler uses the same set of transformations for different pipelining techniques to have a fair comparison. These transformations consist of \textit{dead code elimination} (DCE), \textit{common subexpression elimination} (CSE) or \textit{global value numbering} (GVN), and \textit{partial evaluation} (inlining and constant propagation). These transformations are provided out of the box by DBLAB~\cite{dblablb}, which we use as our testbed. Also, the scalar replacement transformation is always applied unless it is clearly specified. We do not use any data-structure specialization transformations or inverted indices for these experiments. Finally, all experiments use DBLAB's in-memory row-store representation.

For compiling the generated
programs throughout our evaluation we use version 2.9 of the CLang
compiler. 
We use the most aggressive optimization strategy provided by the CLang compiler (the "-O3" optimization flag)\footnote{Although with this optimization flag it takes more time for the optimizing compiler to compile the queries, we observed similar results with the "-O1" optimization flag. This optimization flag provides all the transformation passes used in HyPer~\cite{Neumann11} except global value numbering (GVN). This transformation is not needed in our case, as it is already provided by DBLAB~\cite{dblablb}.}.
Finally, for C data structures we use the GLib library (version 2.42.1).

Our evaluation consists of two parts. First, by using micro benchmarks we demonstrate better the differences between different query engines. Then, for more complex queries we use the TPC-H~\cite{tpch} benchmark. We demonstrate how the different query engines behave in more complicated scenarios.
\subsection{Micro Benchmarks}
\label{sec:exp_micro}
The micro benchmarks belong to three categories: First, queries consisting of only selection and aggregation without group by leading to a single result. Second, queries consisting of selection, projection, sort, and limit operations, which will return a list of results. Finally, queries with selection and different join operators, such as hash join, merge join, and semi hash join, which are followed by an aggregation operator resulting to a single result. All these queries use generated \tpch databases at scaling factor 8, unless specified otherwise. The corresponding SQL queries for all these queries are shown in Table~\ref{tbl:micro_queries}.

\para{Sensitivity to Selectivity} The behavior of different engines for a simple query with one selection operator followed by an aggregation for different selectivities are shown in Figure~\ref{fig:selectivity}. For highly selective queries, the Volcano pull engine is behaving better. This is because the unnecessary elements are skipped faster in the inner tight loop, whereas in the other engines the outer loop is responsible for skipping them. A similar effect was shown in~\cite{pantela2015one} in the context of push engines and vectorized engines. For higher selectivities, in most cases, the push engine performs best. The Visitor-based stream-fusion engine offers almost the same performance as push engine, whereas the scalar-based stream-fusion engine (the one which only uses the scalar replacement transformation and not the Visitor pattern) is worse than other engines in most cases.

\begin{figure}[t]
\centering
\includegraphics[width=\columnwidth]{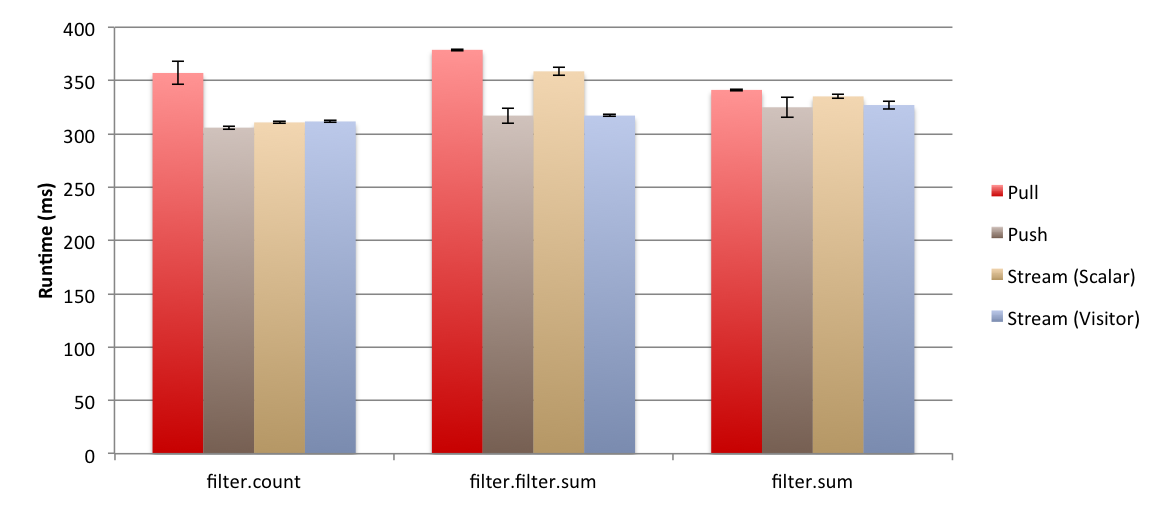}
\caption{Simple queries with aggregated result.}
\figurebottom
\label{fig:simple_agg}
\end{figure}

\para{Aggregated Single Pipeline} Figure~\ref{fig:simple_agg} demonstrates the performance obtained by each engine for queries with a single pipeline which produce a single result by summing over one column. The push engine is behaving slightly better than the pull engine in the presence of only a single filter operation. However, the stream-fusion engines hide this limitation of pull engines.

The difference is more obvious whenever there are chains of selection operations. A similar effect was shown in HyPer~\cite{Neumann11} in the case of using up to four consecutive selection operations. Again the Visitor-based stream-fusion engine is resolving this practical limitation of pull engines. From a practical point of view, as the query optimizer is merging all conjunctive predicates into a single selection operator, this case never happens in practice.

\begin{figure}[t]
\centering
\includegraphics[width=\columnwidth]{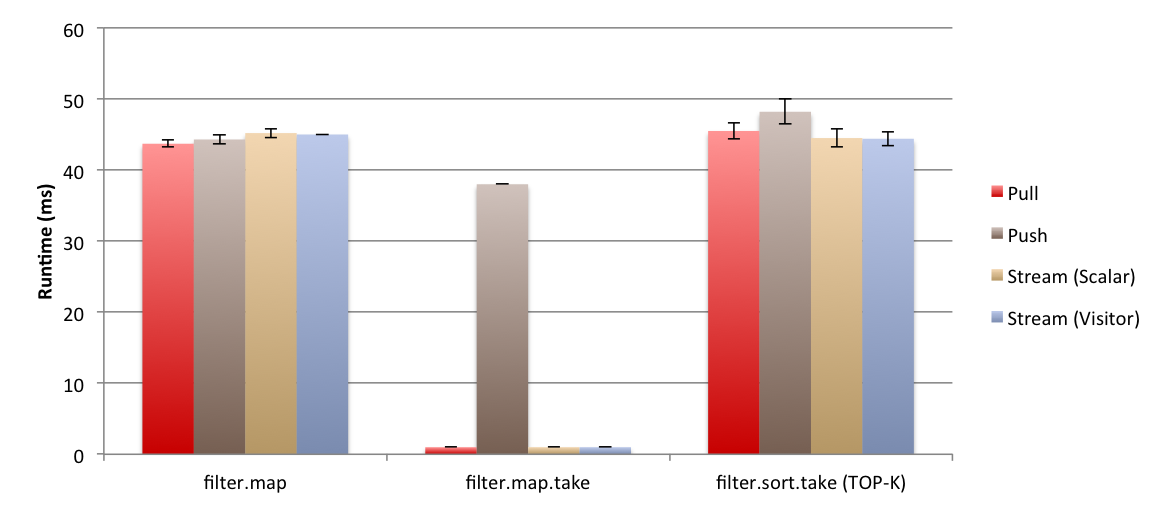}
\caption{Simple queries with a list of results.}
\figurebottom
\label{fig:simple_list}
\end{figure}

\begin{figure}[t]
\centering
\includegraphics[width=\columnwidth]{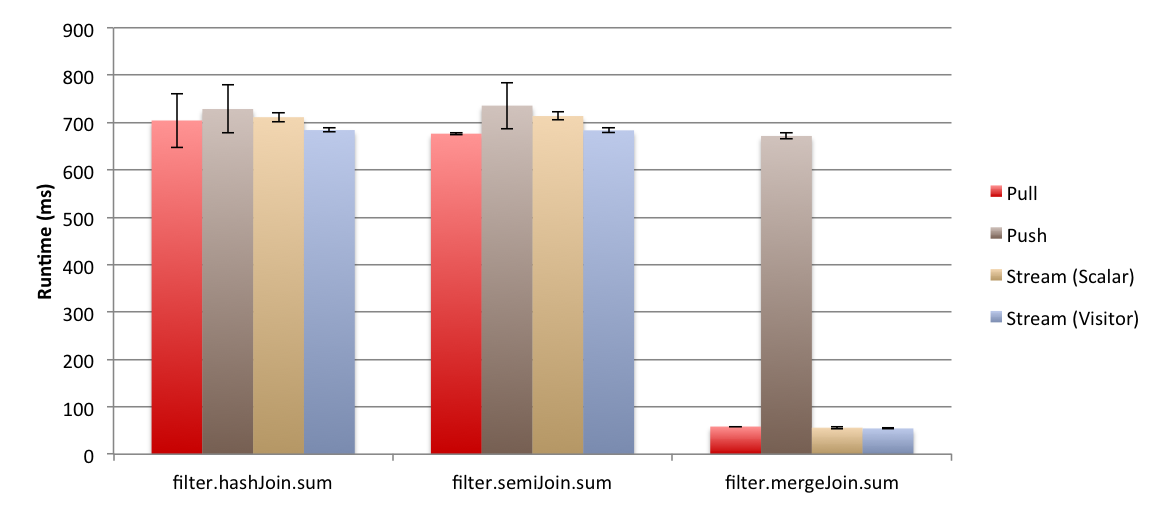}
\caption{Join queries with aggregated result.}
\figurebottom
\label{fig:join_agg}
\end{figure}

\para{Single Pipeline} Figure~\ref{fig:simple_list} shows the results for single pipeline queries which do not contain any aggregation, hence producing a list of elements. For this experiment, we use 1GB of generated data. In the first query, in which a selection is followed by a projection operator, all engines behave similarly. However, in the third query, which returns top-k elements after filtering unnecessary elements, the push engine is performing worse than pull engines. This is because the push engine breaks the pipeline when using the limit query operator (c.f. Section~\ref{sec:engine_push}). This situation is more obvious in the second query which consists of a selection, a projection, and a limit operator. In this case the pull engines do not require traversing all the elements and can stop immediately after reaching the limit. However, the push engine should wait until all elements are produced to be able to finish the execution. A similar phenomenon has been observed for pull-based and push-based fusion techniques for Java 8 streaming API in~\cite{biboudis2015streams}.

\para{Aggregated Single Join} Finally, we investigate the performance of different join operations, which is demonstrated in Figure~\ref{fig:join_agg}. In the case of hash join and semi hash join operators, there is no obvious difference among the engines. However, in the case of merge join, there is a great advantage for pull engines in comparison with the push engine. This is mainly because the push engine cannot pipeline both inputs of a merge join. Hence, it is forced to break the pipeline in one of the inputs (c.f. Section~\ref{sec:engine_push})\footnote{The stream-fusion engine should have a special care for handling merge joins followed by filter operations. By skipping the elements in the main loop of merging, many CPU cycles are wasted for retrieving the next satifying element. However, accessing them by using a similar approach to the Iterator model (keep iterating until the next satisfying element is found in a tight loop) gives a better performance.}.

\begin{figure}[t]
\centering
\includegraphics[width=\columnwidth]{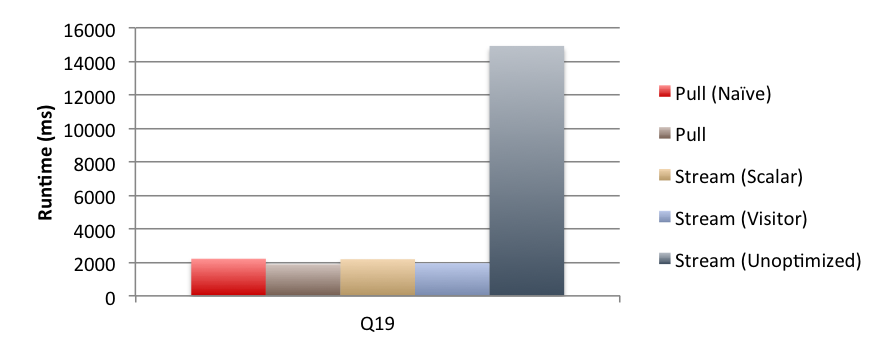}
\caption{The performance comparison of several variants of pull-based engines on \tpch query 19.}
\figurebottom
\label{fig:smart_pull}
\end{figure}

\subsection{Macro Benchmarks}
In this section, we investigate scenarios which are happening more often in practice. To do so, we use the larger and more complicated analytical queries defined in the \tpch benchmark. First, we show the impact of fine-grained optimizations as well as our \smartpull{} way of implementing pull engines on one of \tpch queries. Then, we investigate the impact of different engines on 12 \tpch queries. All these experiments use 8 GBs of \tpch generated data.

\begin{figure*}[t]
\centering
\includegraphics[width=\textwidth]{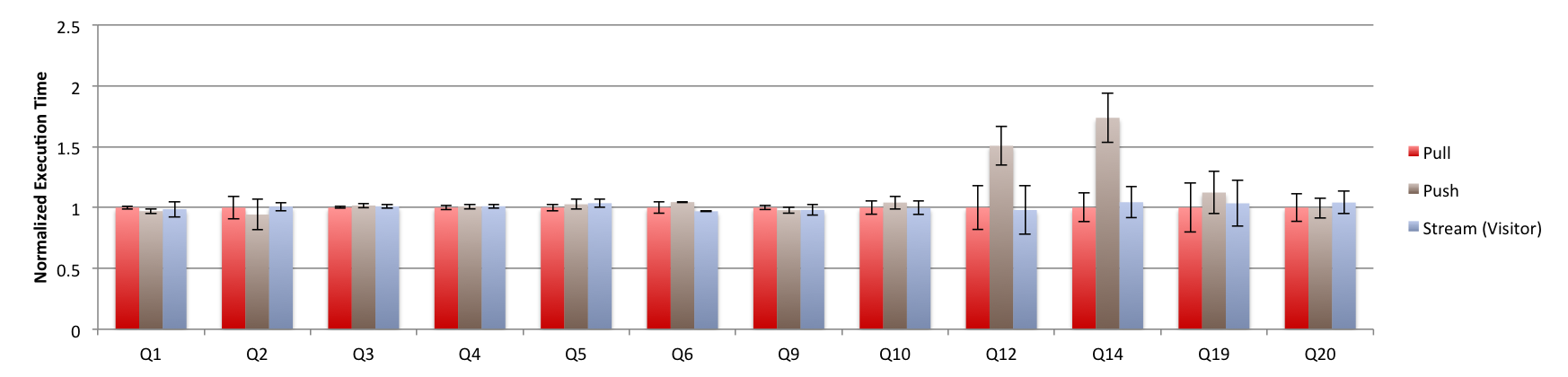}
\caption{Performance of different query engines for TPCH queries.}
\figurebottom
\label{fig:tpch_perf}
\end{figure*}

\para{\SmartPull{} Pull Engine Implementation} A na\"ive implementation of the selection operator in a pull-based query engine, invokes the \code{next} method of its source operator twice. This can exponentially grow the code size in the case of chain of selection operators. This case is not frequent in practice, since the selection operator is mainly used rightly after the scan operator. However, in the case of \tpch query 19 the selection operator is used after a join\footnote{An alternative implementation is to fuse the selections happening after joins in the join operator itself. The experiments performed in~\cite{schuhexperimental} are based on this assumption for join operators. This means that the join operator is not a pure join operator, but a super operator containing a join operator followed by a selection operator. For the purposes of this paper we do not consider such cases.}. Figure~\ref{fig:smart_pull} shows that the \smartpull{} implementation of the selection operator in pull engines, which is shown using the ``Pull'' label throughout this section, improves performance by 15\%. One of the main reasons is that the \smartpull{} implementation generates around 40\% less query processing code in comparison with the na\"ive implementation for query processing in these two queries. This improves instruction cache locality as a larger part of the code can fit into the instruction cache.

\para{Removing Intermediate Object Allocations} Figure~\ref{fig:smart_pull} shows that heap allocating intermediate \code{Step} objects degrades performance in an order of magnitude. Also, the Visitor pattern for \code{Step} objects improves performance by 50\% in comparison with the case in which heap allocations are converted to stack allocations.
Furthermore, our experiments show that for \tpch queries converting heap allocations to stack allocations (either by Visitor pattern or scalar replacement) decreases the memory consumption from 14 GBs to 11 GBs.

\para{Different Engines on Analytical Queries} Figure~\ref{fig:tpch_perf} shows the performance of several \tpch queries using different engines. Overall, this figure shows that the difference between engines is not in terms of "orders of magnitude"; in most cases, improvements are minor. This is because the comparison is performed in a fair scenario in which specialization is performed on all engines, in contrast with previous work in which operator inlining was not applied to pull engines~\cite{legobase}. 

The benchmarked queries can be divided into the following two categories.
The first category consists of the queries which are performing almost equally in all engines. In some cases, we see a minor improvement in push engines mainly because of the better control-flow of the generated code which allows the underlying compiler to generate better machine code. However, even in such cases the difference is marginal.

The second category consists of the queries which perform better in pull engines. This is mainly because of using merge join and limit operators. Query 14 falls into this category because of its use of the limit operator. This query has an average 80\% speed up for a pull engine in comparison with a push engine. Also, query 12 uses the merge join operator and has an average 70\% speed up in comparison with a push-based query engine. It is important to note that in query 12, the query plan that uses a merge join is almost two times faster than the one that uses hash join. This is because both input relations are already sorted on the join key. Hence, the merge join implementation can perform the join on the fly, as opposed to the hash join implementation which needs to construct an intermediate hash table while joining two input relations.

The stream-fusion engine always uses the Visitor pattern throughout this experiment. Interestingly, it is performing as well as push engines, whenever control flow is important. Furthermore, in the cases where push engines require to break the pipeline (the limit and merge join operators) the stream-fusion engine is performing as well as pull engines. This makes the stream-fusion engine an appropriate choice for query engines. 

\section{Discussion: Parallelism}
\label{sec:related}
In this section, we discuss how the results of this paper apply to parallel query engines. {\em Intra-operator parallelism} is one of the main ways of achieving parallelism in query engines. In this approach, data is {\em split} among different threads and each thread is responsible for performing the computation on its associated chunk in a sequential manner. At the end, the results computed by different threads are {\em merged} into a single result.

The split and merge operators~\cite{mehta1995managing} are injected between pipeline breaker operators (such as aggregation and hash join operators). Hence, each thread is sequentially computing the result of a chain of pipelining operators. As a result, by using the same split and merge operators, the only difference between pull and push-based engines is how efficiently they compute the result of a chain of piplining operators. Hence, given a fair environment for pull and push-based engines for intra-operator parallelism, the experimental results we show in this paper for single-threaded scenarios can be expected to match those for multi-threaded scenarios.

One of the key decisions for intra-operator parallelism is the time when the partitioning decision is made. If this decision is made during query compilation time, it is called {\em plan-driven}. If making this decision is postponed until the runtime, it is called {\em morsel-driven}~\cite{Leis:2014:MPN:2588555.2610507}. The key advantage of the latter one is using the runtime information and performing better load balancing by using work-stealing~\cite{blumofe1999scheduling}. However, the partitioning decision choice is also independent of the type of query engine. All types of query engines can use both approaches, and the impact of work-stealing scheduling is similar on both of them.

Similar efforts have been conducted in the PL community for parallelism in collection programs. As an example, morsel-driven parallelism~\cite{Leis:2014:MPN:2588555.2610507} shares similar ideas, such as work-stealing scheduling, with parallel collection programming libraries~\cite{prokopec2011generic}.

\section{Conclusion}

If one effects a fair comparison of push and pull-based query processing -- particularly if one attempts to inline code in both approaches as much as possible -- neither approach clearly outperforms the other. We have discussed the reasons for this, and indeed, when considered closely how each approach fundamentally works, it should seem rather surprising if either approach dominated the other performance-wise.

We have also drawn close connections to three fundamental approaches to loop fusion in programming languages -- fold, unfold, and stream fusion. As it turns out, there is a close analogy between pull engines and unfold fusion on one hand and push engines and fold fusion on the other.

Finally, we have applied the lessons learned about the weaknesses of either approach and propose a new approach to building query engines which draws its inspiration from stream fusion and combines the individual advantages of pull and push engines, avoiding their weaknesses.

\begin{small}
\bibliographystyle{abbrv}
\bibliography{references} 

\begin{thebibliography}{10}

\bibitem{streambase}
{StreamBase Systems}, {http://www.streambase.com}.

\bibitem{DBLP:journals/pvldb/AhmadK09}
Y.~Ahmad and C.~Koch.
\newblock {DBToaster}: {A} {SQL} compiler for high-performance delta processing
  in main-memory databases.
\newblock {\em {PVLDB}}, 2(2):1566--1569, 2009.

\bibitem{spark-sql}
M.~Armbrust, R.~S. Xin, C.~Lian, Y.~Huai, D.~Liu, J.~K. Bradley, X.~Meng,
  T.~Kaftan, M.~J. Franklin, A.~Ghodsi, and M.~Zaharia.
\newblock {Spark SQL: Relational Data Processing in Spark}.
\newblock SIGMOD '15, pages 1383--1394, New York, NY, USA, 2015. ACM.

\bibitem{biboudis2015streams}
A.~Biboudis, N.~Palladinos, G.~Fourtounis, and Y.~Smaragdakis.
\newblock Streams {\`a} la carte: Extensible pipelines with object algebras.
\newblock In {\em 29th European Conference on Object-Oriented Programming},
  page 591, 2015.

\bibitem{blumofe1999scheduling}
R.~D. Blumofe and C.~E. Leiserson.
\newblock Scheduling multithreaded computations by work stealing.
\newblock {\em Journal of the ACM (JACM)}, 46(5):720--748, 1999.

\bibitem{bohm1985automatic}
C.~B{\"o}hm and A.~Berarducci.
\newblock Automatic synthesis of typed $\lambda$-programs on term algebras.
\newblock {\em Theoretical Computer Science}, 39:135--154, 1985.

\bibitem{monad-calc-1}
V.~Breazu-Tannen, P.~Buneman, and L.~Wong.
\newblock {\em Naturally embedded query languages}.
\newblock Springer, 1992.

\bibitem{monad-calc-2}
V.~Breazu-Tannen and R.~Subrahmanyam.
\newblock {\em Logical and computational aspects of programming with
  sets/bags/lists}.
\newblock Springer, 1991.

\bibitem{Buchlovsky2006309}
P.~Buchlovsky and H.~Thielecke.
\newblock A type-theoretic reconstruction of the visitor pattern.
\newblock {\em Electronic Notes in Theoretical Computer Science}, 155:309 --
  329, 2006.

\bibitem{choi1999escape}
J.-D. Choi, M.~Gupta, M.~Serrano, V.~C. Sreedhar, and S.~Midkiff.
\newblock Escape analysis for java.
\newblock {\em Acm Sigplan Notices}, 34(10):1--19, 1999.

\bibitem{Coutts07streamfusion}
D.~Coutts, R.~Leshchinskiy, and D.~Stewart.
\newblock Stream fusion. from lists to streams to nothing at all.
\newblock In {\em ICFP '07}, 2007.

\bibitem{crotty2015tupleware}
A.~Crotty, A.~Galakatos, K.~Dursun, T.~Kraska, U.~{\c{C}}etintemel, and S.~B.
  Zdonik.
\newblock Tupleware:" big" data, big analytics, small clusters.
\newblock In {\em CIDR}, 2015.

\bibitem{Diaconu:2013:HSS:2463676.2463710}
C.~Diaconu, C.~Freedman, E.~Ismert, P.-A. Larson, P.~Mittal, R.~Stonecipher,
  N.~Verma, and M.~Zwilling.
\newblock Hekaton: Sql server's memory-optimized oltp engine.
\newblock In {\em Proceedings of the 2013 ACM SIGMOD International Conference
  on Management of Data}, SIGMOD '13, pages 1243--1254, New York, NY, USA,
  2013. ACM.

\bibitem{Emir:2007:MOP:2394758.2394779}
B.~Emir, M.~Odersky, and J.~Williams.
\newblock Matching objects with patterns.
\newblock ECOOP'07, pages 273--298, Berlin, Heidelberg, 2007. Springer-Verlag.

\bibitem{monoid-comprehension}
L.~Fegaras and D.~Maier.
\newblock Optimizing object queries using an effective calculus.
\newblock {\em ACM Trans. Database Syst.}, 25(4):457--516, Dec. 2000.

\bibitem{gedik-sigmod:08}
B.~Gedik, H.~Andrade, K.-L. Wu, P.~Yu, and M.~Doo.
\newblock {SPADE}: the {System S} declarative stream processing engine.
\newblock In {\em SIGMOD}, 2008.

\bibitem{gibbons2009essence}
J.~Gibbons and B.~C. d.~S. Oliveira.
\newblock The essence of the iterator pattern.
\newblock {\em Journal of Functional Programming}, 19(3-4):377--402, 2009.

\bibitem{foldr-fusion-1}
A.~Gill, J.~Launchbury, and S.~L. Peyton~Jones.
\newblock A short cut to defore- station.
\newblock FPCA, pages 223--232. ACM, 1993.

\bibitem{Volcano}
G.~Graefe.
\newblock {Volcano-an extensible and parallel query evaluation system}.
\newblock {\em IEEE Transactions on Knowledge and Data Engineering},
  6(1):120--135, 1994.

\bibitem{ferry-2}
T.~Grust, M.~Mayr, J.~Rittinger, and T.~Schreiber.
\newblock {FERRY:} database-supported program execution.
\newblock SIGMOD 2009, pages 1063--1066. ACM.

\bibitem{ferry-1}
T.~Grust, J.~Rittinger, and T.~Schreiber.
\newblock Avalanche-safe {LINQ} compilation.
\newblock {\em PVLDB}, 3(1-2):162--172, Sept. 2010.

\bibitem{query-comprehension}
T.~Grust and M.~Scholl.
\newblock How to comprehend queries functionally.
\newblock {\em Journal of Intelligent Information Systems}, 12(2-3):191--218,
  1999.

\bibitem{Hinze:2010:TPF:2050135.2050137}
R.~Hinze, T.~Harper, and D.~W.~H. James.
\newblock Theory and practice of fusion.
\newblock In {\em Proceedings of the 22Nd International Conference on
  Implementation and Application of Functional Languages}, IFL'10, pages
  19--37, Berlin, Heidelberg, 2011. Springer-Verlag.

\bibitem{Hirzel:2014:CSP:2597757.2528412}
M.~Hirzel, R.~Soul{\'e}, S.~Schneider, B.~Gedik, and R.~Grimm.
\newblock A catalog of stream processing optimizations.
\newblock {\em ACM Comput. Surv.}, 46(4):46:1--46:34, Mar. 2014.

\bibitem{Hofer:2010:MDL:1868294.1868307}
C.~Hofer and K.~Ostermann.
\newblock Modular domain-specific language components in scala.
\newblock In {\em Proceedings of the Ninth International Conference on
  Generative Programming and Component Engineering}, GPCE '10, pages 83--92,
  New York, NY, USA, 2010. ACM.

\bibitem{jones1993glasgow}
S.~P. Jones, C.~Hall, K.~Hammond, W.~Partain, and P.~Wadler.
\newblock The glasgow haskell compiler: a technical overview.
\newblock In {\em Proc. UK Joint Framework for Information Technology (JFIT)
  Technical Conference}, volume~93. Citeseer, 1993.

\bibitem{fold-based-fusion}
M.~Jonnalagedda and S.~Stucki.
\newblock Fold-based fusion as a library: A generative programming pearl.
\newblock In {\em Proceedings of the 6th ACM SIGPLAN Symposium on Scala}, pages
  41--50. ACM, 2015.

\bibitem{karpathiotakis2016fast}
M.~Karpathiotakis, I.~Alagiannis, and A.~Ailamaki.
\newblock Fast queries over heterogeneous data through engine customization.
\newblock {\em Proceedings of the VLDB Endowment}, 9(12):972--983, 2016.

\bibitem{karpathiotakis2015just}
M.~Karpathiotakis, I.~Alagiannis, T.~Heinis, M.~Branco, and A.~Ailamaki.
\newblock Just-in-time data virtualization: Lightweight data management with
  vida.
\newblock In {\em CIDR}, 2015.

\bibitem{legobase}
Y.~Klonatos, C.~Koch, T.~Rompf, and H.~Chafi.
\newblock Building efficient query engines in a high-level language.
\newblock {\em {PVLDB}}, 7(10):853--864, 2014.

\bibitem{legobase-errata}
Y.~Klonatos, C.~Koch, T.~Rompf, and H.~Chafi.
\newblock Errata for "building efficient query engines in a high-level
  language": Pvldb 7(10):853-864.
\newblock {\em Proc. VLDB Endow.}, 7(13):1784--1784, Aug. 2014.

\bibitem{DBLP:conf/pods/Koch10}
C.~Koch.
\newblock Incremental query evaluation in a ring of databases.
\newblock PODS 2010, pages 87--98. ACM, 2010.

\bibitem{kochmanifesto}
C.~Koch.
\newblock Abstraction without regret in database systems building: a manifesto.
\newblock {\em IEEE Data Eng. Bull.}, 37(1):70--79, 2014.

\bibitem{dbtoaster}
C.~Koch, Y.~Ahmad, O.~Kennedy, M.~Nikolic, A.~N\"otzli, D.~Lupei, and
  A.~Shaikhha.
\newblock {DBToaster}: higher-order delta processing for dynamic, frequently
  fresh views.
\newblock {\em VLDBJ}, 23(2):253--278, 2014.

\bibitem{krikellas}
K.~Krikellas, S.~Viglas, and M.~Cintra.
\newblock {Generating code for holistic query evaluation}.
\newblock In {\em ICDE}, pages 613--624, 2010.

\bibitem{Leis:2014:MPN:2588555.2610507}
V.~Leis, P.~Boncz, A.~Kemper, and T.~Neumann.
\newblock {Morsel-driven Parallelism: A NUMA-aware Query Evaluation Framework
  for the Many-core Age}.
\newblock SIGMOD '14, pages 743--754, New York, NY, USA, 2014. ACM.

\bibitem{lorie1974xrm}
R.~A. Lorie.
\newblock {\em XRM: An extended (N-ary) relational memory}.
\newblock IBM, 1974.

\bibitem{mehta1995managing}
M.~Mehta and D.~J. DeWitt.
\newblock Managing intra-operator parallelism in parallel database systems.
\newblock In {\em VLDB}, volume~95, pages 382--394, 1995.

\bibitem{linq}
E.~Meijer, B.~Beckman, and G.~Bierman.
\newblock {LINQ: Reconciling Object, Relations and XML in the .NET Framework}.
\newblock SIGMOD '06, pages 706--706. ACM, 2006.

\bibitem{Murray:2011:SAO:1993498.1993513}
D.~G. Murray, M.~Isard, and Y.~Yu.
\newblock Steno: Automatic optimization of declarative queries.
\newblock PLDI '11, pages 121--131, New York, NY, USA, 2011. ACM.

\bibitem{Nagel:2014:CGE:2732977.2732984}
F.~Nagel, G.~Bierman, and S.~D. Viglas.
\newblock Code generation for efficient query processing in managed runtimes.
\newblock {\em PVLDB}, 7(12):1095--1106.

\bibitem{Neumann11}
T.~Neumann.
\newblock Efficiently {C}ompiling {E}fficient {Q}uery {P}lans for {M}odern
  {H}ardware.
\newblock {\em PVLDB}, 4(9):539--550, 2011.

\bibitem{pantela2015one}
S.~Pantela and S.~Idreos.
\newblock One loop does not fit all.
\newblock In {\em Proceedings of the 2015 ACM SIGMOD International Conference
  on Management of Data}, pages 2073--2074. ACM, 2015.

\bibitem{PG88}
J.~Paredaens and D.~V. Gucht.
\newblock Possibilities and limitations of using flat operators in nested
  algebra expressions.
\newblock In {\em Proceedings of the Seventh {ACM} {SIGACT-SIGMOD-SIGART}
  Symposium on Principles of Database Systems, March 21-23, 1988, Austin,
  Texas, {USA}}, pages 29--38, 1988.

\bibitem{pierce2002types}
B.~C. Pierce.
\newblock {\em Types and programming languages}.
\newblock MIT press, 2002.

\bibitem{prokopec2011generic}
A.~Prokopec, P.~Bagwell, T.~Rompf, and M.~Odersky.
\newblock A generic parallel collection framework.
\newblock In {\em Euro-Par 2011 Parallel Processing}, pages 136--147. Springer,
  2011.

\bibitem{schuhexperimental}
S.~Schuh, X.~Chen, and J.~Dittrich.
\newblock An experimental comparison of thirteen relational equi-joins in main
  memory.
\newblock 2016.

\bibitem{dblablb}
A.~Shaikhha, Y.~Klonatos, L.~Parreaux, L.~Brown, M.~Dashti, and C.~Koch.
\newblock How to architect a query compiler.
\newblock SIGMOD'16, 2016.

\bibitem{coroutine-fusion}
O.~Shivers and M.~Might.
\newblock Continuations and transducer composition.
\newblock PLDI '06, pages 295--307. ACM, 2006.

\bibitem{Svenningsson:2002:SFA:581478.581491}
J.~Svenningsson.
\newblock Shortcut fusion for accumulating parameters \& zip-like functions.
\newblock ICFP '02, pages 124--132. ACM, 2002.

\bibitem{tpch}
{Transaction Processing Performance Council}.
\newblock {TPC-H, a decision support benchmark.}
\newblock \url{http://www.tpc.org/tpch}.

\bibitem{query-comprehension-2}
P.~Trinder.
\newblock {Comprehensions, a Query Notation for DBPLs}.
\newblock In {\em Proc. of the 3rd DBPL workshop}, DBPL3, pages 55--68, San
  Francisco, CA, USA, 1992. Morgan Kaufmann Publishers Inc.

\bibitem{DBLP:journals/debu/ViglasBN14}
S.~Viglas, G.~M. Bierman, and F.~Nagel.
\newblock {Processing Declarative Queries Through Generating Imperative Code in
  Managed Runtimes}.
\newblock {\em {IEEE} Data Eng. Bull.}, 37(1):12--21, 2014.

\bibitem{vlissides1995design}
J.~Vlissides, R.~Helm, R.~Johnson, and E.~Gamma.
\newblock Design patterns: Elements of reusable object-oriented software.
\newblock {\em Reading: Addison-Wesley}, 49(120):11, 1995.

\bibitem{deforestation}
P.~Wadler.
\newblock Deforestation: Transforming programs to eliminate trees.
\newblock In {\em ESOP'88}, pages 344--358. Springer, 1988.

\bibitem{monad-comprehension}
P.~Wadler.
\newblock Comprehending monads.
\newblock In {\em Proceedings of the 1990 ACM Conference on LISP and Functional
  Programming}, LFP '90, pages 61--78, New York, NY, USA, 1990. ACM.

\bibitem{rdd}
M.~Zaharia, M.~Chowdhury, T.~Das, A.~Dave, J.~Ma, M.~McCauley, M.~J. Franklin,
  S.~Shenker, and I.~Stoica.
\newblock {Resilient Distributed Datasets: A Fault-tolerant Abstraction for
  In-memory Cluster Computing}.
\newblock NSDI'12. USENIX Association.

\end{thebibliography}
\end{small}

\appendix
\section{Micro Benchmark Queries}

\begin{table}[h]
\newcommand{\tblheader}[1]{\hspace{0.1cm}\textbf{#1:}}
\centering
\begin{tabular}{| l | l | l | }
\hline
\tblheader{filter.count}  & \tblheader{filter.sum} & \tblheader{filter.filter.sum} \\ 
\begin{lstlisting}[language=SQL]
SELECT COUNT(*) 
FROM LINEITEM 
WHERE L_SHIPDATE >=
 DATE '1995-12-01'
\end{lstlisting}
&
\begin{lstlisting}[language=SQL]
SELECT 
 SUM(L_DISCOUNT
 * L_EXTENDEDPRICE) 
FROM LINEITEM 
WHERE L_SHIPDATE >=
 DATE '1995-12-01'
\end{lstlisting}
& 
\begin{lstlisting}[language=SQL]
SELECT SUM(L_DISCOUNT
 * L_EXTENDEDPRICE) 
FROM LINEITEM 
WHERE (L_SHIPDATE >= 
 DATE '1995-12-01')
 AND (L_SHIPDATE <
 DATE '1997-01-01')
\end{lstlisting}\\ \hline
\tblheader{filter.map}  & \tblheader{filter.sort.take} & \tblheader{filter.map.take} \\ 
\begin{lstlisting}[language=SQL]
SELECT L_DISCOUNT
 * L_EXTENDEDPRICE 
FROM LINEITEM 
WHERE L_SHIPDATE >=
 DATE '1995-12-01'
\end{lstlisting}
&
\begin{lstlisting}[language=SQL]
SELECT 
  L_EXTENDEDPRICE
FROM LINEITEM 
WHERE L_SHIPDATE >= 
 DATE '1995-12-01'
ORDER BY L_ORDERKEY
LIMIT 1000
\end{lstlisting}
& 
\begin{lstlisting}[language=SQL]
SELECT L_DISCOUNT
 * L_EXTENDEDPRICE 
FROM LINEITEM 
WHERE L_SHIPDATE >=
 DATE '1995-12-01'
LIMIT 1000
\end{lstlisting}\\ \hline
\multicolumn{3}{|c|}{\tblheader{filter.XJoin(filter).sum}} \\ 
\multicolumn{3}{|}{}
\begin{lstlisting}[language=SQL]
 SELECT SUM(O_TOTALPRICE) FROM LINEITEM, ORDERS
 WHERE O_ORDERDATE >= DATE '1998-11-01' 
   AND L_SHIPDATE >= DATE '1998-11-01'
   AND  O_ORDERKEY = L_ORDERKEY
\end{lstlisting}
\hfill\vline \\ \hline
\end{tabular}
\caption{SQL queries of micro benchmark queries.}
\label{tbl:micro_queries}
\end{table}
\end{document}